\documentclass[10pt,journal,compsoc]{IEEEtran}
\usepackage{amsmath}
\usepackage{amssymb}
\usepackage{mathtools,amsthm}
\usepackage{comment}
\usepackage{graphicx}
\usepackage{caption}
\usepackage{subcaption}
\usepackage{hyperref}
\usepackage{xcolor}

\theoremstyle{definition}

\usepackage[ruled,linesnumbered]{algorithm2e}
\usepackage{enumitem}
\usepackage{booktabs}

\ifCLASSOPTIONcompsoc
  \usepackage[nocompress]{cite}
\else
  \usepackage{cite}
\fi

\begin{document}

\title{Learning to Incentivize: LLM-Empowered Contract for AIGC Offloading in Teleoperation}

\markboth{Journal of \LaTeX\ Class Files,~Vol.~14, No.~8, August~2015}%
{Shell \MakeLowercase{\textit{et al.}}: Bare Demo of IEEEtran.cls for Computer Society Journals}

\author{Zijun~Zhan,
        Yaxian~Dong,
        Daniel~Mawunyo~Doe,
        Yuqing~Hu,
        Shuai~Li,
        Shaohua Cao,
        and~Zhu~Han,
        \IEEEcompsocitemizethanks{\IEEEcompsocthanksitem{Zijun Zhan is with the Department of Electrical and Computer Engineering, University of Houston, 4800 Calhoun Rd, Houston, TX 77004. E-mail: zzhan@uh.edu}
         \IEEEcompsocthanksitem{Yaxian Dong and Yuqing Hu are with the Department of Architectural Engineering, The Pennsylvania State University, University Park, PA 16802, USA (E-mail: yzd5221@psu.edu and yfh5204@psu.edu)}
	\IEEEcompsocthanksitem{Daniel Mawunyo Doe is with the Department of Electrical and Computer Engineering, Prairie View A\&M University, 100 University Dr, Prairie View, TX 77446. Email: dmdoe@pvamu.edu}
         \IEEEcompsocthanksitem{Shuai Li is with the Department of Civil \& Coastal Engineering, University of Florida, Gainesville, FL 32611. E-mail: shuai.li@ufl.edu}
         \IEEEcompsocthanksitem{Shaohua Cao is with the Qingdao Institute of Software, College of Computer Science and Technology, China University of Petroleum (East China), Qingdao 266580, China. E-mail:shaohuacao@upc.edu.cn}
         \IEEEcompsocthanksitem{Zhu Han is with the Department of Electrical and Computer Engineering, University of Houston, 4800 Calhoun Rd, Houston, TX 77004, and also with the Department of Computer Science and Engineering, Kyung Hee University, Seoul, South Korea, 446-701. E-mail: hanzhu22@gmail.com}}
}

\IEEEtitleabstractindextext{
\begin{abstract}
With the rapid growth in demand for AI-generated content (AIGC), edge AIGC service providers (ASPs) have become indispensable. However, designing incentive mechanisms that motivate ASPs to deliver high-quality AIGC services remains a challenge, especially in the presence of information asymmetry. In this paper, we address bonus design between a teleoperator and an edge ASP when the teleoperator cannot observe the ASP’s private settings and chosen actions (diffusion steps). We formulate this as an online learning contract design problem and decompose it into two subproblems: ASP’s settings inference and contract derivation. To tackle the NP-hard setting-inference subproblem with unknown variable sizes, we introduce a large language model (LLM)-empowered framework that iteratively refines a naive seed solver using the LLM’s domain expertise. Upon obtaining the solution from the LLM-evolved solver, we directly address the contract derivation problem using convex optimization techniques and obtain a near-optimal contract. Simulation results on our Unity-based teleoperation platform show that our method boosts the teleoperator’s utility by $5 \sim 40\%$ compared to benchmarks, while preserving positive incentives for the ASP. The code is available at \url{https://github.com/Zijun0819/llm4contract}.
\end{abstract}

}

\maketitle

\IEEEdisplaynontitleabstractindextext

\IEEEpeerreviewmaketitle


\IEEEraisesectionheading{\section{Introduction} \label{sec:1}}
\IEEEPARstart{T}{he} remarkable performance of AI-generated content (AIGC) services is driving their market growth exponentially. As per the report in \cite{aigcmarket}, the AIGC market will increase from \$1.6B in 2023 to \$18.7B in 2030, with an annual growth rate of 25.39\%. To underpin this substantial market, the AIGC services are gradually burgeoning in the domains of education \cite{chen2024empowering}, healthcare \cite{chen2024revolution}, and teleoperation \cite{zhan2024vision}, etc. In this paper, we will dig into a specific case of applying AIGC services in teleoperation. Concretely, to ensure seamless teleoperation in dark areas, such as tunnels, drainage systems, and nighttime construction sites, a generative diffusion model can restore normal-light images from low-light conditions \cite{zhan2024vision, zhan2025distributionally}.

The latency requirement for teleoperation ranges from 100 to 300 ms \cite{kamtam2024network}. AIGC services are generally computationally intensive, and on-site devices may not have sufficient computational resources. To this end, many scholars proposed the edge AIGC framework \cite{du2024diffusion, du2024enabling, li2024multi}, in which the AIGC service provider (ASP) supplies AIGC services in the edge server. The edge server is not only proximal to end users but also has ample computational resources \cite{zhou2024potential, chen2025revenue}. The introduction of ASP can help reduce the latency of AIGC-empowered teleoperation. Moreover, a well-designed incentive mechanism in edge AIGC networks can enable sustainable and high-quality AIGC service provision \cite{wen2024diffusion}.

Numerous scholars proposed various incentive mechanisms for edge AIGC services \cite{zhan2024vision,zhan2025distributionally,wen2024diffusion,wen2023freshness,wen2024learning,ye2024optimizing}. We categorize them into ASP-centric and user-centric (teleoperator-centric) types. For the ASP-centric incentive mechanism, its primary objective is to maximize the utility of the ASP while ensuring the rationality of user incentives. In contrast, the user-centric incentive mechanism aims to maximize user utility while ensuring the rationality of the ASP's incentives. In this paper, we focus on designing user-centric incentive mechanisms for edge AIGC networks, so as to ensure user engagement. It is worth noting that information asymmetry may exist during the design of incentive mechanisms in edge AIGC networks \cite{zhan2024vision, zhan2025distributionally, wen2024diffusion}. For instance, users have limited information regarding the available computational resources of the ASP.

A mature tool that can tackle the information asymmetry within the incentive mechanism design is contract theory \cite{li2023book}. Specifically, from the perspective of teleoperators, we can formulate a series of contract bundles, (Latency, Reward), with properties of incentive compatibility and incentive rationality. The ASP will select the contract bundle according to its available computational resources, thereby maximizing its utility. The teleoperator rewards the ASP only when it delivers the AIGC service within the required latency. We can formulate an effective incentive mechanism via contract theory in terms of the AIGC service latency. However, if we shift the lens to AIGC service quality, it may be a challenge for us to formulate an incentive mechanism.

\begin{figure}[!t]
    \centering
    \includegraphics[width=0.85\linewidth]{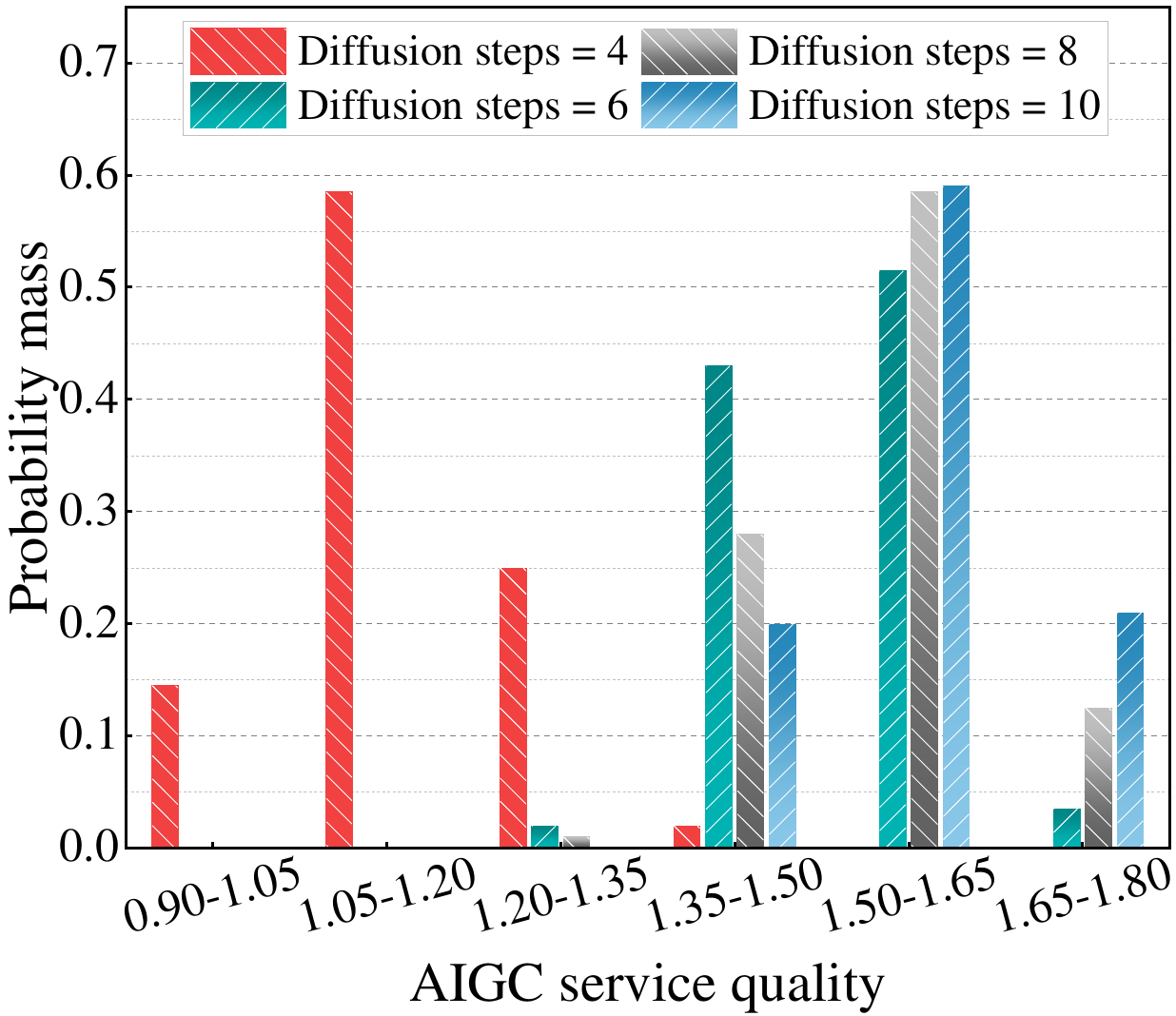}
    \caption{PMF of the AIGC services quality under varying diffusion steps, in which the AIGC service quality is the sum of LPIPS \cite{zhang2018unreasonable} and SSIM \cite{wang2004image} of processed AIGC results.}
    \label{fig1}
    \vspace{-5mm}
\end{figure}

\begin{figure*}[!t]
    \centering
    \includegraphics[width=1\linewidth]{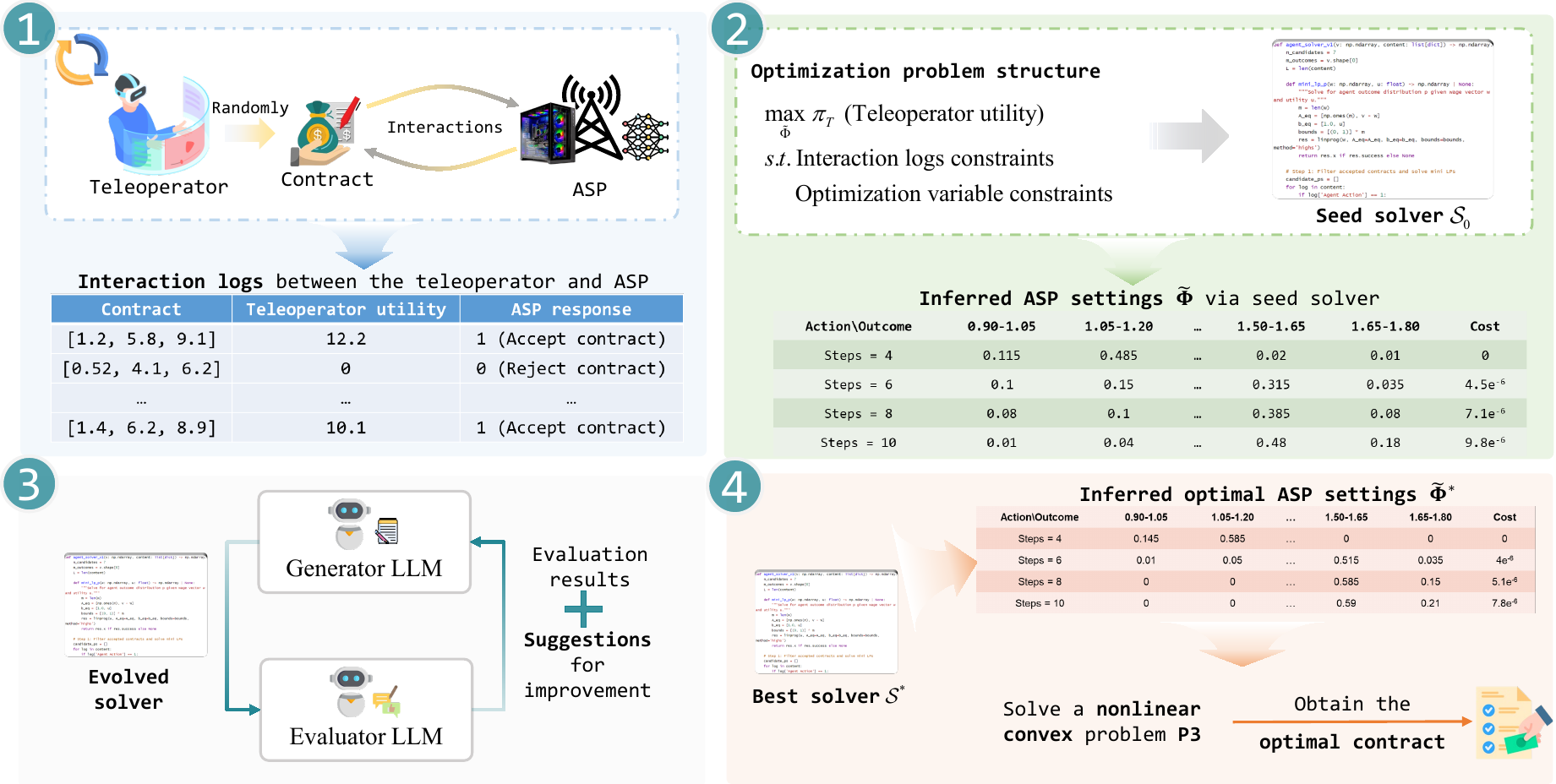}
    \caption{Framework of LLM-empowered online learning contract theory for deriving the near-optimal bonus (contract) regarding the AIGC services quality.}
    \label{fig2}
    \vspace{-5mm}
\end{figure*}

The hidden information in the latency case is the type of ASPs; in contrast, the hidden information in the AIGC service quality case is the actions of ASPs. Concretely, as the results depicted in Fig. \ref{fig1}, when the ASP undertakes different diffusion steps (actions), the distribution of the AIGC services quality varies. In previous works \cite{li2021contract, doe2023promoting, liu2024prosecutor}, they modeled this type of contract theory problem as a moral hazard problem. However, they assume the teleoperator (principal) knows the ASP's (agent) setting (i.e., the mapping of diffusion steps to the AIGC service quality distribution and the cost of varying diffusion steps) in advance. Without this unrealistic assumption, we need to model the incentive mechanism design for AIGC service quality as an online learning contract theory problem, which is APX-hard \cite{ho2014adaptive, dutting2019simple, bacchiocchi2023learning, guruganesh2024contracting, wang2023deep}. Therefore, we formulate the research question in this paper as follows: \emph{How can we design an incentive mechanism that elicits high-quality AIGC services, thereby maximizing the teleoperator's utility under the hidden action of the edge ASP?}

The essence of the research question is an online learning contract design problem, but the off-the-shelf solution rarely exists. The majority of works \cite{ho2014adaptive, dutting2019simple, bacchiocchi2023learning, guruganesh2024contracting} focus on the theoretical level, aiming to prove the lower regret bound for the online learning contract design problem. A recent work \cite{wang2023deep} proposes a deep learning–based solution for the general online learning contract theory problem; however, it requires thousands of training samples. In other words, we can address the research question only after thousands of contract interactions have occurred between the teleoperator and the edge ASP. To this end, we proposed an LLM-empowered solution to effectively solve the research question, inspired by recent works that utilize LLMs to design extraordinary heuristics for solving NP-hard problems \cite{liu2024evolution, ye2024reevo, yao2025multi}.

We present our proposed LLM-empowered online learning contract theory in Fig. \ref{fig2}, which includes four steps. Firstly, we will randomly generate contracts and observe the interaction results between the teleoperator and the edge ASP. After 100 rounds, we can obtain a series of interaction tuples (Contract, Teleoperator utility, ASP response). Subsequently, upon the historical interaction tuples and the optimization problem structure, we will design a naive seed solver to infer the edge ASP's setting. Next, we harness the power of LLM to iteratively refine the seed solver with the aid of multiple LLM agents. Lastly, we select the solver with the highest evaluation score during the self-evolution to infer the optimal edge ASP setting. Based on the inferred edge ASP setting, we derive a near-optimal contract by using convex optimization techniques to solve a traditional moral hazard problem (P3 in Section \ref{sec:3.4}). In summary, we conclude the contributions of this paper as follows:
\begin{enumerate}
    \item [1.] We propose a framework of LLM-empowered incentive mechanism design under information asymmetry. With our proposed framework, the teleoperator can derive a near-optimal bonus (contract) under the hidden action of edge ASP, which can elicit the edge ASP to take higher diffusion steps and provide high-quality AIGC services. In this framework, we utilize the LLM to refine a solver for inferring the edge ASP's setting.

    \item [2.] We extend the traditional moral hazard problem by eliminating the assumption that the principal (teleoperator) knows the agent's (edge ASP) setting in advance. Concretely, we model the moral hazard problem as an online learning contract design problem, in which the teleoperator needs to learn edge ASP's settings via historical interaction logs prior to deriving the contract.

    \item [3.] We propose utilizing the LLM to address the online learning contract design problem. Due to the coupled variables in the online learning contract design problem, we decompose it into two subproblems: the edge ASP setting inference problem and the contract derivation problem. The edge ASP's setting inferring problem is APX-hard; we leverage the LLM to evolve a solver for it. Upon acquiring the solution from the LLM-evolved solver, we directly solve the contract derivation problem using convex optimization techniques.

    \item [4.] We conduct numerous experiments to demonstrate the scalability, sensitivity, and effectiveness of our proposed LLM-empowered framework. The experimental results demonstrate that our proposed method can augment the teleoperator utility by around $5 \sim 40\%$ in comparison with benchmarks under almost all experimental configurations. Moreover, our proposed method ensures a positive incentive to the edge ASP.
\end{enumerate}

We organize the rest of this paper as follows. Section \ref{sec:2} provides a systematic literature review. Section \ref{sec:3} presents the system model and formulates the optimization problem. In Section \ref{sec:4}, we illustrate the details of how we leverage the LLM to derive the near-optimal contract, along with the analysis of algorithms. Section \ref{sec:5} systematically evaluates the LLM-empowered solution in terms of scalability, effectiveness, and sensitivity. Finally, Section \ref{sec:6} concludes the paper.

\section{Background and Related Works} \label{sec:2}

\subsection{Background of Online Learning Contract Theory} \label{sec:2.1}
Contract theory examines the design of incentive mechanisms between two parties: a principal, who offers a contract, and an agent, who selects actions in response to the provided incentives. Contract design problems typically arise under conditions of information asymmetry, wherein the agent possesses private information that is not accessible to the principal \cite{hart1986theory, stiglitz1975theory, holmstrom1979moral}. Two canonical forms of information asymmetry extensively studied in contract theory are adverse selection and moral hazard \cite{bolton2004}.

Adverse selection arises when the agent possesses private information about their inherent characteristics or "type" prior to entering into a contract \cite{liu2018incentive, yu2024contract}. Such characteristics may include skill level, reliability, cost structure, or risk profile—factors that are not directly observable by the principal \cite{bolton2004, wang2020consortium, wang2022motilearn}. Because the principal cannot distinguish between high-quality and low-quality agents ex ante, contracts intended for high-quality agents may inadvertently attract low-quality agents, resulting in suboptimal outcomes. To mitigate adverse selection, the principal typically designs a menu of contracts (i.e., a screening mechanism) that encourages each agent type to self-select the contract best aligned with their private information. This approach ensures incentive compatibility and can partially reveal agent types through their contract choices \cite{zhang2017contract}.

Moral hazard arises after a contract is signed, when the agent takes actions that influence the outcome but these actions are not fully observable or contractible by the principal \cite{zhang2017mh}. This scenario presents a challenge, as the agent may act in their own interest rather than that of the principal, particularly when such actions are costly or effort-intensive. To address moral hazard, contracts are typically designed to align the agent’s incentives with those of the principal, often by linking rewards or penalties to observable outcomes that correlate with the agent’s effort \cite{bolton2004}. Traditional approaches to the moral hazard problem assume that the principal knows the functional relationship between the agent’s actions and the resulting outcomes, as well as the agent’s utility function—including the cost associated with different actions \cite{zhang2017mh, li2021contract, doe2023promoting, liu2024prosecutor, tian2021contract, ding2017moral, zhang2017multi, li2025optimal, ismail2024enhancing}.

Online learning contract theory extends traditional moral hazard models by considering more practical scenarios in which the principal initially lacks complete knowledge of the agent's private parameters \cite{ho2014adaptive, dutting2024algorithmic}. These private parameters include both the mapping between the agent's actions and the outcomes observed by the principal, as well as the agent's cost of undertaking different actions. In this context, the principal must iteratively learn and infer these hidden parameters through repeated interactions before effectively addressing the moral hazard problem. The seminal work by Ho et al. \cite{ho2014adaptive} introduced the concept of online learning contract theory by modeling the moral hazard problem as a repeated principal–agent interaction in crowdsourcing markets. Specifically, the principal offers a contract to a randomly arriving, strategic worker, who selects an effort level (unobservable to the principal) that determines both output quality and personal cost. The principal observes only the output, and updates the contract in each round of interaction to maximize long-term utility.

\begin{table*}[htbp]
\centering 
\caption{Comparison Between Our Work and Previous Works.}
\label{Table:1}
\begin{tabular}{p{4.5cm} p{2cm} p{2.5cm} p{1.8cm} p{2cm} p{2.3cm}}
\toprule
Ref.           & AIGC networks & Adverse selection & Moral hazard & Agent settings & Sample efficiency \\ \midrule
\cite{zhan2024vision, wen2023freshness, wen2024diffusion, wen2024learning, ye2024optimizing, zhan2025distributionally}   & $\checkmark$                     & $\checkmark$                           & $\times$                             & \text{N/A}                  & \text{N/A}    \\
\cite{li2021contract, doe2023promoting, tian2021contract}        & $\times$                        & $\checkmark$                           & $\checkmark$                             & $\checkmark$                  & \text{N/A} \\
\cite{liu2024prosecutor}        & $\checkmark$                        & $\times$                           & $\checkmark$                             & $\checkmark$                  & \text{N/A}  \\
\cite{ding2017moral, zhang2017multi, li2025optimal, ismail2024enhancing}   & $\times$                        & $\times$                           & $\checkmark$                             & $\checkmark$                  & \text{N/A}  \\
\cite{ho2014adaptive, zhu2022sample, cohen2022learning, bacchiocchi2025learning, dutting2019simple, bacchiocchi2023learning, wang2023deep} & $\times$                        & $\times$                           & $\checkmark$                             & $\times$                  & $\times$ \\
\midrule
\textbf{Our work}        & $\checkmark$                        & $\times$                           & $\checkmark$                             & $\times$                  & $\checkmark$  \\ \bottomrule
\end{tabular}
\vspace{1mm}
\\
\footnotesize
\textcolor{gray}{$^*$ \text{N/A} indicates not applicable, and the column of Agent settings means whether the principal can access the agent's private settings.}
\vspace{-5mm}
\end{table*}

\subsection{Related Works} \label{sec:2.2}
Given the research question in this paper, we will review the recent literature from three perspectives. In Section \ref{sec:2.2.1}, we review recent works on utilizing contract theory to design incentive mechanisms in edge AIGC networks. Next, we revisit recent works that utilize contract theory to address hidden actions during the design of incentive mechanisms in Section \ref{sec:2.2.2}. Finally, we present recent works regarding online learning contract design in Section \ref{sec:2.2.3}.

\subsubsection{Contract Theory in Edge AIGC Networks} \label{sec:2.2.1}
With the maturity of AIGC services, the authors in \cite{du2023exploring} proposed an edge AIGC services offloading framework, enabling mobile users to access AIGC services seamlessly. To maintain sustainable and high-quality AIGC services, several works \cite{zhan2024vision, wen2023freshness, wen2024diffusion, wen2024learning, ye2024optimizing, zhan2025distributionally} have recognized the necessity of an incentive mechanism in the edge AIGC services offloading framework. Meanwhile, in these works, they also observed that information asymmetry prevails in incentive mechanism design for edge AIGC networks. Consequently, they proposed different incentive mechanisms for varying edge AIGC networks by utilizing contract theory.

The authors in \cite{zhan2024vision} observed that edge servers cannot observe the difficulty distributions of AIGC tasks in advance. Therefore, they fused a vision–language model with contract theory to classify AIGC task difficulty on the fly and priced a differentiated AIGC task offloading. The authors in \cite{wen2023freshness} developed an Age of Information (AoI)-driven contract that rewards UAVs for uploading fresh data to support AIGC model training. They maximized base-station utility while curing hidden-cost asymmetry stemming from UAV-private sensing costs and update frequencies. Similarly, the authors in \cite{wen2024learning} proposed a Proximal Policy Optimization (PPO)-based contract theory to derive the optimal contract, thereby addressing the issue that the data collection cost of mobile devices is hidden information to ASPs.

Given that ASPs' AIGC model complexity is hidden from users, and users may possess risk-biased valuations, the authors in \cite{wen2024diffusion} combined contract theory with prospect theory to generate user-centric contracts. The authors in \cite{ye2024optimizing} introduced a contract theory-based two-stage incentive mechanism for AIGC task allocation. In this work, the users' subjective gain per unit of AIGC service quality is hidden from the ASP. The authors in \cite{zhan2025distributionally} devised a Wasserstein distributionally robust optimization (DRO) contract to derive robust latency-reward contract bundles for teleoperators. The design addresses the information asymmetry that teleoperators cannot observe the available capacity of ASPs, as well as ensuring that the fluctuation of AIGC service quality will not affect the teleoperators' utility.

These studies examined the integration of contract theory into edge AIGC networks from various perspectives, laying a solid foundation for future research. However, these works focus on the design of incentive mechanisms under hidden-type information asymmetry. They did not consider the contract theory-based incentive mechanism design from the perspective of hidden actions. Our work will investigate how to design an effective incentive mechanism to address the hidden action in edge AIGC networks.

\subsubsection{Hidden Action Contract Design} \label{sec:2.2.2}
The previous works that apply contract theory to design incentive mechanisms in edge AIGC networks focus on the hidden type aspect, i.e., the adverse selection problem. In this section, we will review several recent works that focus on addressing the moral hazard problem, thereby deriving incentive mechanisms under hidden action.

The authors in \cite{li2021contract} considered that the resource devoted by a validator for transaction verification in a blockchain system is a hidden action to the blockchain beacon chain. To this end, they proposed a bonus mechanism based on the framework of contract theory to incentivize validators to devote more resources to transaction verification. Analogously, the authors in \cite{doe2023promoting} designed a bonus mechanism to motivate blockchain users to exert more resources in maintaining the full node of the blockchain. The authors in \cite{liu2024prosecutor} observed that ASPs might intentionally reduce the resources they invest (such as computation power) in generating AI outputs, thereby representing a moral hazard. The authors deployed contract theory to optimize the payment schemes between clients and ASPs.

The authors in \cite{tian2021contract} modeled federated learning (FL) training as a moral hazard problem, as the training effort exerted by FL clients was unobservable to the FL server. In \cite{ding2017moral}, contract theory was applied to address moral hazard in multi-hop cooperative communications, where the cooperative user's willingness to contribute energy for content transmission constituted the hidden action. The work in \cite{zhang2017multi} introduced a multi-dimensional moral hazard model for mobile crowdsourcing systems, where mobile workers’ efforts spanned multiple metrics such as accuracy, timeliness, and coverage; the authors utilized contract theory to derive the optimal incentive scheme. In \cite{li2025optimal}, the moral hazard problem was formulated around the hidden effort of edge caching nodes in disseminating short videos, which remained unobservable to service providers. Finally, the authors in \cite{ismail2024enhancing} considered a scenario in which virtual service providers could reduce service quality after receiving compensation, due to unobservable post-contract actions.

Similar to \cite{liu2024prosecutor}, we also consider the moral hazard problem that exists in edge AIGC networks. As depicted in Fig. \ref{fig1}, the diffusion step of the generative diffusion model undertaken by ASPs is hidden from teleoperators, and the action of ASPs will impact the utility of teleoperators. Diverging from the previous work \cite{li2021contract, doe2023promoting, liu2024prosecutor, tian2021contract, ding2017moral, zhang2017multi, li2025optimal, ismail2024enhancing}, we eliminate the unrealistic assumption that the teleoperator knows the edge ASP's setting.

\subsubsection{Online Learning Contract Design} \label{sec:2.2.3}
When we do not know the edge ASPs' settings—the mapping of diffusion steps to the AIGC service quality distribution and the cost of actions—the hidden action contract design will shift to online learning contract design \cite{dutting2024algorithmic}. In this way, we need to iteratively learn ASP's settings before addressing the moral hazard problem and then derive the optimal incentive mechanism (contract). Subsequently, we will review several representative works in this emerging field.

The authors in \cite{ho2014adaptive} first proposed the online learning contract design problem and modeled it as a multi-armed bandit problem. Analogously, the authors in \cite{zhu2022sample} modeled the problem as a continuum-armed bandit problem and used the upper confidence bound to solve it. The authors in \cite{cohen2022learning} also modeled a continuous contract space online learning contract design problem. However, they discretized the contract space into a finite but rich menu of candidate contracts and utilized tools from multi-armed bandit theory to derive the optimal contract. The authors in \cite{bacchiocchi2025learning} extended the work of \cite{zhu2022sample} by focusing on the setting of agents with small action space and proposed a Discover-and-Cover algorithm to derive the near-optimal contract. The authors in \cite{dutting2019simple} assumed the principal (teleoperator) has moment-based information of the agent (ASP) settings, and proposed to use robust optimization to derive the optimal contract. The authors in \cite{bacchiocchi2023learning} considered a case where the agent's (ASP) action space is small, and proposed a discover-and-cover algorithm to find the optimal contract iteratively. The authors in \cite{wang2023deep} observed that the principal utility is discontinuous with respect to the contract. Therefore, they proposed a discontinuous neural network to model the principal utility function and utilized gradient ascent to find the optimal contract.

The aforementioned works are pioneering work in online learning contract design. However, one primary issue is that the previous works require substantial contract interaction logs between ASPs and teleoperators before deriving the optimal contract. In this paper, we unleash the power of LLM to design an effective online learning contract solver, thereby deriving a near-optimal incentive mechanism with minimal interaction logs. To the best of our knowledge, this is the first work to utilize LLM in designing incentive mechanisms under information asymmetry. Additionally, we summarized the comparison between our work and current works in Table \ref{Table:1} for clarity.

\section{System Model and Problem Formulation} \label{sec:3}
In this section, we commence with the introduction of the background and key notations in Section \ref{sec:3.1}. Next, we present the utility model of the edge ASP and teleoperator in Sections \ref{sec:3.2} and \ref{sec:3.3}. Lastly, we formulate the optimization problem in Section \ref{sec:3.4}.

\subsection{System Model and Notations} \label{sec:3.1}
In this paper, we consider a scenario where a teleoperator company needs to purchase generative diffusion-based AIGC services from an edge ASP to achieve seamless teleoperation working in low-light areas. During the provision of AIGC services, latency and quality are two crucial metrics. Previous works \cite{zhan2025distributionally, wen2024diffusion} have leveraged contract theory to formulate contracts in a bundle of (Latency, Reward) to motivate the ASP to meet the latency requirement. Analogously, we can use contract theory to formulate a bonus incentive mechanism \cite{li2021contract} in the form of (Quality, Bonus), thereby incentivizing ASP to take larger diffusion steps for high-quality AIGC services provision. However, the unknown edge ASPs' setting $\Phi = (\mathbf{P},\mathbf{c})$ hinders the derivation of an optimal bonus incentive mechanism (contract) for the teleoperator. Here, $\mathbf{P}$ is the mapping between diffusion steps and the AIGC service quality distribution. $\mathbf{c}$ is the additional cost of the edge ASP when it takes different actions.

We model the contract derivation as an online learning contract design problem. By referring  to \cite{wang2023deep}, we define the problem with elements $\mathcal{C}=(\mathcal{D}, \mathcal{R}, \mathcal{O}, \mathcal{A}, \Phi, q)$. Since $\Phi$ is unobservable to the teleoperator, we need to leverage the historical contract interaction logs $\mathcal{D}$ between the teleoperator and the edge ASP to infer edge ASP's setting $\tilde{\Phi} = (\tilde{\mathbf{P}},\tilde{\mathbf{c}})$. Here, we define $\mathcal{D}=\{(\mathbf{r}_k, \pi_T(\mathbf{r}_k), \mathbb{I}(\mathbf{r}_k))|k \in [K]\}$, in which $[K]=\{1,\cdots,k,\cdots,K\}$. Regarding the $k$-th interaction log, the teleoperator commences with randomly generate a contract $\mathbf{r}=\{r_m|m\in[M]\} \in \mathcal{R} \subset \mathbb{R}_{\geq 0}^{M}$ over the finite and discrete outcome space $\mathcal{O}=\{o_m|m \in [M]\}$. Next, the edge ASP selects a diffusion step (action) $a_n$ from the discrete and finite action space $\mathcal{A}$ as per the contract $\mathbf{r}$, in which $\mathcal{A}=\{a_n|n \in [N]\}$. Action $a_n$ leads to an AIGC service quality distribution $\mathbf{p}_n = p(\cdot|a_n)$ over $\mathcal{O}$ and incurs an additional processing cost $c_n \in \mathbb{R}_{\geq 0}$ to the edge ASP. Subsequently, the teleoperator perceives an outcome $o_m$ and obtains the utility of $\pi_T(\mathbf{r})$. Lastly, in light of the teleoperator cannot observe $a_n$, we utilize $\mathbb{I}(\mathbf{r}) = \{-1, 1\}$ to indicate whether the edge ASP rejects or accepts the contract $\mathbf{r}$. For clarity, we consider $\mathbf{P} \in \mathbb{R}^{N \times M}$ and $\mathbf{c} \in \mathbb{R}^{N}$.

In this paper, we consider each outcome $o_m$ to be an AIGC service quality range, defined as $o_m = [l_m, u_m)$. We use the median value of $o_m$, i.e., $\delta_m = (l_m+u_m)/2$, to calculate the valuation of $o_m$ for the teleoperator. The valuation function of $o_m$ is defined as $q: \mathcal{O} \rightarrow \mathbb{R}_{\geq 0}$. After calculation, we utilize $\mathbf{q}=\{q_m|m \in [M]\}$ to represent the utility of teleoperator over $\mathcal{O}$.

\subsection{Utility Model of Edge ASP} \label{sec:3.2}
We assume the utility of the edge ASP stems from two parts: the utility of the contract and the utility of the AIGC services subscription. Upon this, we define the utility function of the edge ASP as:
\begin{equation} \label{eq:1}
    \pi_A(a_i;\mathbf{r},\Phi)=\underbrace{\mathbb{E}_{o_m \sim p(\cdot|a_n)}{[r_m]} - c_n}_{\Delta_b} + \underbrace{r_s-c_t}_{\Delta_s}.
\end{equation}
Here, $\Delta_b$ indicates the utility of the contract. $r_s$ is the subscription fee paid by the teleoperator for accessing the AIGC services, $c_t$ is the training cost of the edge ASP for training the AIGC model, and $\Delta_s$ is the utility of the AIGC services subscription. In this paper, we consider a utility model designed per image. It is worth noting that the AIGC service subscription will bundle a default diffusion step $a_1$, and we set $c_1 = 0$.

In each contract interaction, the edge ASP selects the action $a^*(\mathbf{r},\Phi)$ that maximizes its utility as per the contract designed by the teleoperator. Therefore, we define the utility of the edge ASP in each contract interaction as:
\begin{equation} \label{eq:2}
    \pi_A = \max_a \pi_A(a;\mathbf{r},\Phi).
\end{equation}
We term (\ref{eq:2}) as the incentive compatibility (IC) constraint. Additionally, the edge ASP will only accept the contract if
\begin{equation} \label{eq:3}
    \pi_A \geq \Delta_s.
\end{equation}
We refer to (\ref{eq:3}) as the incentive rationality (IR) constraint. We set $a^*(\mathbf{r},\Phi) = a_1$ if the edge ASP reject the contract $\mathbf{r}$. Without other specifications, we will use $a^{*}$ to represent $a^*(\mathbf{r},\Phi)$ in the rest of the paper for clarity.

\subsection{Utility Model of Teleoperator} \label{sec:3.3}
We consider the utility model of the teleoperator to consist of three parts: the gain from the quality of AIGC services, the bonus pay to the edge ASP, and the cost of AIGC services subscription. We assume the teleoperator utility is $0$ if the edge ASP rejects the contract. Therefore, we model the utility function of the teleoperator as:
\begin{equation} \label{eq:4}
    \pi_{T}(\mathbf{r})=
    \begin{cases}
        \mathbb{E}_{o_m \sim p(\cdot|a^{*})}\left[q_m-r_m\right]-r_s, & \text{if } a^{*}\neq a_1,\\
        0, & \text{otherwise},
    \end{cases}
\end{equation}
by referring to \cite{zhan2025distributionally, wen2024diffusion}, we set $q_m$ as
\begin{equation} \label{eq:5}
    q_m = ln(1+\alpha \delta_m).
\end{equation}
Here, $\alpha$ is the coefficient mapping the AIGC services quality to the teleoperator's gain. $\delta_m$ is the median value of $o_m$. Due to the nature of the logarithm function, the teleoperator's valuation regarding the AIGC service quality has a rapid initial growth rate that gradually diminishes as $\delta_m$ increases.

\subsection{Optimization Problem Formulation} \label{sec:3.4}
The optimization problem in this pape r is maximizing the teleoperator utility without access to $\Phi$. We define the optimization problem as a standard moral hazard form:
\begin{subequations}
    \begin{align}
        P1: ~ \max_{\mathbf{r}} \quad &\mathbb{E}_{o_m \sim p(\cdot|a^{*})}\left[q_m-r_m\right]-r_s, \label{eq:6a} \\
        s.t. \quad &a^* = \arg \max_{a} \pi_A(a;\mathbf{r},\Phi) \label{eq:6b} \\
        &\pi_A \geq \Delta_s. \label{eq:6c}
    \end{align}
\end{subequations}
The difficulty in addressing P1 lies in the condition that we cannot directly solve the IC in (\ref{eq:6b}) and IR in (\ref{eq:6c}) constraints. In other words, we can tackle P1 with ease if we know $\Phi$, since then P1 is a convex optimization problem. Therefore, we reformulate P1 into two sub-problems, P2 and P3, via the historical contract interaction logs $\mathcal{D}$. Concretely, we formulate P2 as a $\Phi$ approximation problem, which requires us to find and approximate a feasible $\tilde{\Phi}$ via $\mathcal{D}$. We define P2 as a feasibility problem as follows:
\begin{subequations}
    \begin{align}
        P2: ~ find \quad &(\mathbf{\mathbf{X}}, \tilde{\Phi}), \label{eq:7a} \\
        s.t. \quad &\sum_{n=1}^{N} x_{n,k} = 1, \forall~k, \label{eq:7b} \\
        & \sum_{m=1}^{M} \tilde{p}_{n,m} = 1, \forall~n, \label{eq:7c}\\
        & \pi_T(\mathbf{r}_k) \leq {\tilde{\mathbf{p}}_n}^T(\mathbf{q} - \mathbf{r}_k) - r_s + L(1-x_{n,k}), \nonumber \\ & \forall~n,k, \label{eq:7d}\\
        & \pi_T(\mathbf{r}_k) \geq {\tilde{\mathbf{p}}_n}^T(\mathbf{q} - \mathbf{r}_k) - r_s - L(1-x_{n,k}), \nonumber \\ & \forall~n,k, \label{eq:7e}\\
        & \pi_A(a_n;\mathbf{r}_k, \tilde{\Phi}) \geq \Delta_s - L(1-x_{n,k}), \forall~n,k, \label{eq:7f}\\
        & \pi_A(a_n;\mathbf{r}_k, \tilde{\Phi}) \geq \pi_A(a_{n^{\prime}};\mathbf{r}_k) - L(1-x_{n,k}), \nonumber \\ & \forall~n\neq n^{\prime}, \forall~n,k, \label{eq:7g}\\
        & \pi_A(a_n;\mathbf{r}_k) \leq \Delta_s, \forall~\mathbb{I}(\mathbf{r}_k) = 0, \label{eq:7h} \\
        & x_{n,k} \in \{0,1\}, \tilde{p}_{n,m} \in [0,1], \tilde{c}_n \in [0, L]. \label{eq:7i}
    \end{align}
\end{subequations}
Here,  Eq. (\ref{eq:7b}) indicates the selection constraint of the edge ASP, indicates the edge ASP selects action $a_n$ in the $k$-th interaction regarding the contract $\mathbf{r}_k$. Eq. (\ref{eq:7c}) is the probability constraint, which means for each action $a_n$ the sum of the probabilities over the outcome space $\mathcal{O}$ equals 1. Eqs. (\ref{eq:7d}) and (\ref{eq:7e}) is the utility matching constraint, indicates the teleoperator utility under edge ASP's action $a_n$, contract $\mathbf{r}_k$, and the probability mapping vector $\mathbf{p}_n$ under the action $a_n$ should match the real teleoperator utility $\pi_T(\mathbf{r}_k)$. Eqs. (\ref{eq:7f}) and (\ref{eq:7g}) is the IR and IC constraints when edge ASP accepts the contract $\mathbf{r}_k$. Eq. (\ref{eq:7h}) is the contract rejection constraint, which means the IR constraint is violated when the teleoperator formulates the contract $\mathbf{r}_k$. Eq. (\ref{eq:7i}) is the range constraint of variables $\mathbf{X}$, $\tilde{\mathbf{P}}$, and $\tilde{\mathbf{c}}$. $L$ is a large constant used to aid the mathematical formulation.

\begin{figure*}[!t]
    \centering
    \includegraphics[width=1\linewidth]{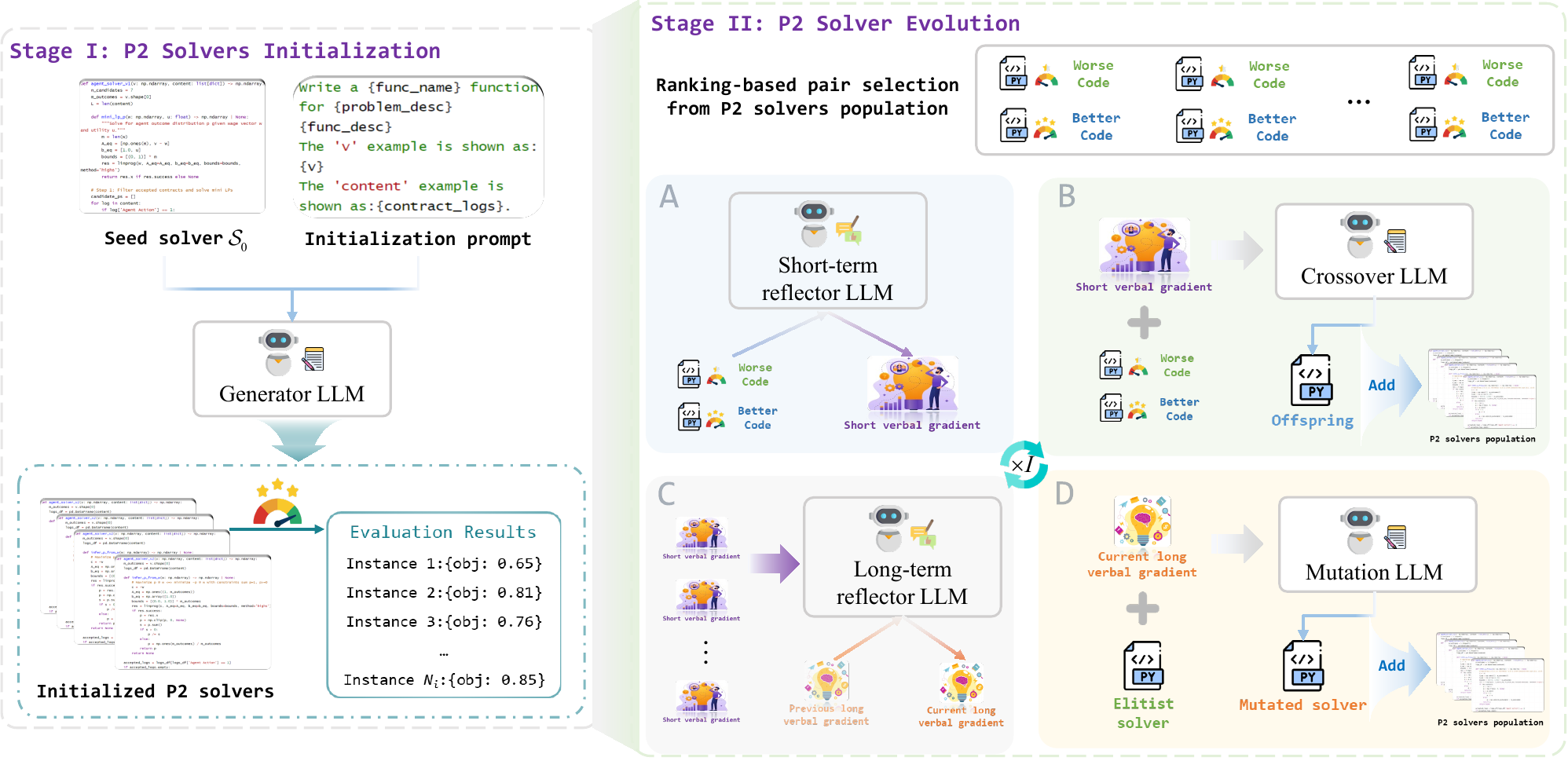}
    \caption{LLM-empowered P2 solver evolution framework, which presents the details of step three in Fig. \ref{fig2}. We present the description of stages I and II in Sections \ref{sec:4.1} and \ref{sec:4.2}, respectively. We present all prompts of LLM agents in Appendix A of the supplementary material.}
    \label{fig3}
    \vspace{-5mm}
\end{figure*}

Notably, it remains a challenge for us to solve P2 directly from twofold. Firstly, P2 is a mixed integer nonlinear programming (MINLP) problem, which is NP-hard. Secondly, we cannot access the size of $N$, which means the dimension of the optimization variables $\mathbf{X}$ and $\tilde{\Phi}$ are uncertain. Therefore, similar to recent online learning contract theory works \cite{ho2014adaptive, dutting2019simple, bacchiocchi2023learning, guruganesh2024contracting, wang2023deep}, we should strive to design an algorithm that derives an approximate solution as accurately as possible.

Once we tackle P2 and acquire $\tilde{\Phi}$, We can formulate P3 in a similar manner as P1:
\begin{subequations}
    \begin{align}
        P3: ~ \arg \max_{\mathbf{\tilde{r}}} \quad &\mathbb{E}_{o_m \sim \tilde{p}(\cdot|a^{*}(\mathbf{\tilde{r}},\tilde{\Phi}))}\left[q_m-\tilde{r}_m \right]-r_s, \label{eq:8a} \\
        s.t. \quad &a^*(\mathbf{\tilde{r}},\tilde{\Phi}) = \arg \max_{a} \pi_A(a;\mathbf{\tilde{r}},\tilde{\Phi}), \label{eq:8b} \\
        &\pi_A \geq \Delta_s. \label{eq:8c}
    \end{align}
\end{subequations}
P3 is a nonlinear convex problem, and we can use convex optimization techniques to tackle it, thereby acquiring the approximated contract $\tilde{\mathbf{r}}$. Regarding the optimization problems formulated in this section, we proposed an LLM-empowered solution and presented the overall framework diagram in Fig. \ref{fig2}. In the next section, we will dig into the details of how we unleash the power of LLM to tackle the optimization problems.

\section{LLM-Empowered Solution} \label{sec:4}

With the problem reformulation in Section \ref{sec:3.4}, the objective function will shift to maximize $\pi_T(\tilde{\mathbf{r}})$ by solving P2 and P3 progressively, and the key is solving P2. With recent exciting advancements in leveraging LLM to solve NP-hard problems \cite{liu2024evolution, ye2024reevo, yao2025multi}, we consider leveraging LLM to tackle P2 effectively. We present the overall framework of LLM-empowered P2 solver in Fig. \ref{fig3}, which includes P2 solvers initialization and P2 solver evolution in two stages. Concretely, in this section, we illustrate the initialization of P2 solvers in Section \ref{sec:4.1} and present the evolution of the P2 solver in Section \ref{sec:4.2}. Lastly, we present and analyze the algorithm of LLM-empowered P2 solver evolution in Section \ref{sec:4.3}.

\subsection{P2 solvers initialization} \label{sec:4.1}

As per the left panel of Fig. \ref{fig3}, the P2 solvers initialization is composed of three major components: seed solver, initialization prompt, and solvers evaluation.
\subsubsection{Seed Solver} \label{sec:4.1.1}
We consider building the seed solver of P2 in a naive manner and specify it in Algorithm \ref{alg:1}.
\begin{enumerate}
    \item [1.] We commence with initializing the parameters in lines 1-3.
    \item [2.] We screen accepted contracts and calculate the possible probability mapping vector $\mathbf{p}_k$ in lines 4-9.
    \item [3.] We aggregate $\Psi_p$ into $\tilde{N}$ vectors via KMeans \cite{sinaga2020unsupervised} and obtain $\tilde{\mathbf{P}}$ in line 10, in which the vector in $\tilde{\mathbf{P}}$ is in ascending order.
    \item [4.] We leverage the IC and IR constraints to approximate the cost vector $\tilde{\mathbf{c}}$ of edge ASP in lines 12-23. When the edge teleoperator accepts the contract, we utilize both IC and IR constraints (\ref{eq:7f}) and (\ref{eq:7g}), in lines 12-20. When the edge teleoperator rejects the contract, we use the IR constraint (\ref{eq:7h}), in lines 21-23.
\end{enumerate}
Notably, we define the P4 in line 6 as:
\begin{subequations}
    \begin{align}
        P4: ~ \min_{\mathbf{p}_k} \quad &{\mathbf{p}_k}^T\mathbf{r}_k, \label{eq:9a} \\
        s.t. \quad &\sum_{m=1}^{M} p_{k,m} = 1, \label{eq:9b} \\
        & {\mathbf{p}_k}^T(\mathbf{q} - \mathbf{r}_k) -r_s = \pi_T(\mathbf{r}_k), \label{eq:9c} \\
        & p_{k,m} \in [0, 1]. \label{eq:9d}
    \end{align}
\end{subequations}
Eq. (\ref{eq:9b}) is the probability constraint, which stems from Eq. (\ref{eq:7c}). Eq. (\ref{eq:9c}) is the teleoperator utility matching constraint, stems from (\ref{eq:7d}) and (\ref{eq:7e}). Eq. (\ref{eq:9d}) is the range constraint, partial of (\ref{eq:7i}). By solving P4, we can acquire a potential probability mapping vector $\mathbf{p}_k$ between the edge ASP's action and the teleoperator's outcome. It is worth noting that we set the objective function as minimizing the reward paid to the edge ASP, which is equivalent to maximizing the teleoperator's utility. We designate the P2 seed solver as $\mathcal{S}_0$ in the rest of the paper.

\subsubsection{Initialization Prompt} \label{sec:4.1.2}
The initialization prompt includes several crucial components: problem description, function description, input instance, and seed function.

\textbf{Problem Description:} We provide a hint to the LLM agent that we need to infer a valid edge ASP setting that satisfies all interaction logs in $\mathcal{D}$ regarding an online learning contract design problem.

\textbf{Function Description:} We describe the input format of the P2 solver and state its requirements, specifying the structure of a valid $\tilde{\Phi}$.

\textbf{Input Instance:} We provide the input illustration of the P2 solver, including the teleoperator's profit over the AIGC service quality $\mathbf{q}$ and the historical interaction logs $\mathcal{D}$.

\textbf{Seed Function:} We provide a runnable Python code, i.e., Algorithm \ref{alg:1}, to the LLM agent for reference.

\subsubsection{Solvers Evaluation} \label{sec:4.1.3}
Upon completion of P2 solvers initialization, we can acquire $N_i$ solvers, denotes as $\{\mathcal{S}_{n_i}|n_i \in [N_i]\}$. We evaluate the fitness of $\mathcal{S}_{n_i}$ by tackling the P3. Concretely, $\mathcal{S}_{n_i}$ is a Python function of the P2 solver with the input of $\mathbf{q}$ and $\mathcal{D}$, and returns a predicted edge ASP's setting $\tilde{\Phi}_{n_i}$. Subsequently, we can use $\tilde{\Phi}_{n_i}$ to solve the P3 and derive a contract $\tilde{\mathbf{r}}_{n_i}$. Lastly, we observe the teleoperator utility $\pi_T(\tilde{\mathbf{r}}_{n_i})$ under contract $\tilde{\mathbf{r}}_{n_i}$ and leverage it as the fitness score of $\mathcal{S}_{n_i}$.

\subsection{P2 solver evolution} \label{sec:4.2}

\begin{algorithm}[!t]
\DontPrintSemicolon
\SetAlgoLined
\KwIn{Teleoperator's profit regarding outcomes $\mathbf{q}$ and historical interaction logs $\mathcal{D}$}
\KwOut{Inferred edge ASP's setting $\tilde{\Phi}=(\tilde{\mathbf{P}},\tilde{\mathbf{c}})$}
\BlankLine
Assume the size of action space $N=\tilde{N}$\;
Initialize candidate probability vector set $\Psi_{p}$\;
Initialize edge ASP's cost vector $\tilde{\mathbf{c}}=0$\;
\ForEach{$(\mathbf{r}_k,\pi_T(\mathbf{r}_k), \mathbb{I}(\mathbf{r}_k)) \in \mathcal{D}$}{
    \If{$\mathbb{I}(\mathbf{r}_k) == 1$}{
        $\mathbf{p}_k \leftarrow$ Solve P4 via $\mathbf{r}_k$ and $\pi_T(\mathbf{r}_k)$\;
        $\Psi_{p} \leftarrow \mathbf{p}_k $\;
    }
}
$\tilde{\mathbf{P}} \leftarrow KMeans(\tilde{N},\Psi_{p})$\;
\ForEach{$(\mathbf{r}_k,\pi_T(\mathbf{r}_k), \mathbb{I}(\mathbf{r}_k)) \in \mathcal{D}$}{
    \If{$\mathbb{I}(\mathbf{r}_k) == 1$}{
        $n_k \leftarrow argmax(\tilde{\mathbf{P}}\mathbf{r}_k)$\;
        \If{$\tilde{\mathbf{c}}[n_k] == 0$}{
            $\tilde{\mathbf{c}}[n_k] = \tilde{\mathbf{P}}[n_k]\mathbf{r}_k)$\;
        }\Else{
            $\tilde{\mathbf{c}}[n_k] = min(\tilde{\mathbf{c}}[n_k], \tilde{\mathbf{P}}[n_k]\mathbf{r}_k))$\;
        }
    }
    \Else{
         $\tilde{\mathbf{c}} = max(\tilde{\mathbf{c}}, \tilde{\mathbf{P}}\mathbf{r}_k)$\;
    }
}
\Return{$\tilde{\Phi}=(\tilde{\mathbf{P}},\tilde{\mathbf{c}})$}\;
\caption{P2 Seed Solver}
\label{alg:1}
\end{algorithm}

\begin{algorithm}[!t]
\DontPrintSemicolon
\SetAlgoLined
\KwIn{Seed solver $\mathcal{S}_0$ and iteration rounds $I$}
\KwOut{Elitist P2 solver $\mathcal{S}^*$}
\BlankLine
\tcc{Initialize and evaluate P2 solvers}
$\{\mathcal{S}_{n_i}| {n_i} \in [N_i]\} \leftarrow GeneratorLLM(\mathcal{S}_0, Init~prompt)$\;
$\{\tilde{{\Phi}}_{n_i}| {n_i} \in [N_i]\} \leftarrow$ Invoke $\{\mathcal{S}_{n_i}| {n_i} \in [N_i]\}$\;
$\{\pi_T(\tilde{\mathbf{r}}_{n_i})| {n_i} \in [N_i]\} \leftarrow$ Tackle P3 via $\tilde{{\Phi}}_{n_i}$ and evaluate $\tilde{\mathbf{r}}_{n_i}$\; 
$\Omega_o \leftarrow \{\mathcal{S}_{n_i}| {n_i} \in [N_i]\}, i = N_i$\;
$\mathcal{S}^* \leftarrow$ Opt the best P2 solver from $\Omega_o$\;
\While{$i \leq I$}{
    \tcc{Begin the epoch of P2 solver evolution}
    $\Omega_n \leftarrow \emptyset$\;
    Rank $\mathcal{S}_s \in \Omega_o$ in a descending order as per $\pi_T(\tilde{\mathbf{r}}_s)$\;
    \ForEach{$n_s \in [N_s/2]$}{
        Opt $\mathcal{S}_b$ and $\mathcal{S}_w$ from $\Omega_o$ with probability proportional to their rank\;
        $\Omega_n \leftarrow \{\mathcal{S}_b, \mathcal{S}_w\}$\;
        $\theta_{n_s} \leftarrow ShortReflectorLLM(\mathcal{S}_b, \mathcal{S}_w)$\;
        $\mathcal{S}_{n_s} \leftarrow CrossoverLLM(\mathcal{S}_b, \mathcal{S}_w, \theta_{n_s})$\;
        $\pi_T(\tilde{\mathbf{r}}_{n_s}) \leftarrow$ Invoke $\mathcal{S}_{n_s}$, tackle P3, and evaluate $\tilde{\mathbf{r}}_{n_s}$\;
        $\Omega_n \leftarrow \{\mathcal{S}_{n_s}\}$\;
        $i += 1$
    }
    $\Theta \leftarrow ShortReflectorLLM(\{\theta_{n_s}| n_s \in [N_s/2]\}, \Theta^{\prime})$\;
    \ForEach{$n_m \in [N_m]$}{
        $\mathcal{S}_{n_m} \leftarrow MutationLLM(\Theta, \mathcal{S}^*)$\;
        $\pi_T(\tilde{\mathbf{r}}_{n_m}) \leftarrow$ Invoke $\mathcal{S}_{n_m}$, tackle P3, and evaluate $\tilde{\mathbf{r}}_{n_m}$\;
        $\Omega_n \leftarrow \{\mathcal{S}_{n_m}\}$\;
        $i += 1$
    }
    $\mathcal{S}^* \leftarrow$ Opt the best P2 solver from $\Omega_n \cup \{\mathcal{S}^*\}$\;
    $\Omega_o \leftarrow \Omega_n$\;
}
\Return{$\mathcal{S}^*$}\;
\caption{LLM-Empowered P2 Solver Evolution}
\label{alg:2}
\end{algorithm}

As per the right panel of Fig. \ref{fig3}, the P2 solver evolution comprises one premise and four steps. The premise is that we need to do a ranking-based pair selection from the old P2 solver population $\Omega_o$. Next, we execute the A-D steps in a manner akin to a genetic algorithm, including short-term reflection, crossover, long-term reflection, and mutation. Notably, the P2 solver evolution will run $I$ iterations and return the elitist P2 solver $\mathcal{S}^{*}$ during the evolution as the final output. In this paper, we consider that one epoch is running the premise and steps A-D, which includes multiple iterations.

\subsubsection{Premise Step} \label{sec:4.2.1}
At the beginning of an epoch, we will sort all solvers in $\Omega_o$ by their fitness score and repeatedly select two P2 solvers, $\mathcal{S}_b$ and $\mathcal{S}_w$, with probability proportional to their rank. Here, $\mathcal{S}_b$ and $\mathcal{S}_w$ represent a better and a worse P2 solver, respectively. We will append $\mathcal{S}_b$ and $\mathcal{S}_w$ to the new P2 solver population $\Omega_n$, and the selection ends when the size of $\Omega_n$ reaches $N_s$. Remarkably, it is crucial to ensure that the fitness scores of $\mathcal{S}_b$ and $\mathcal{S}_w$ are distinct, thereby eliciting the ``short verbal gradient" in the next step.

\subsubsection{Short-Term Reflection (Step A)} \label{sec:4.2.2}
Regarding each $\mathcal{S}_b$ and $\mathcal{S}_w$ selected in the premise step, we will use a short-term reflector LLM agent to generate a ``short verbal gradient" as insight for P2 solver refinement. Concretely, the reflector LLM will compare and analyze $\mathcal{S}_b$ and $\mathcal{S}_w$, then output less than 20 words as ``short verbal gradient" $\theta_{n_s}$. We present one instance of the ``short verbal gradient" as follows: ``\textit{Incorporate LP for p inference, adaptive clustering with silhouette, simplex projection, and strict IR/IC cost adjustments.}" In light of we have opted $N_s$ P2 solvers in the premise step, we can obtain $N_s/2$ ``short verbal gradient", denotes as $\{\theta_{n_s} | n_s \in [N_s/2]\}$.

\subsubsection{Crossover (Step B)} \label{sec:4.2.3}
With $\mathcal{S}_b$ and $\mathcal{S}_w$ and the associated ``short verbal gradient" $\theta_{n_s}$ as input, the crossover LLM agent will generate an offspring $\mathcal{S}_{n_s}$. Similarly, the crossover step will generate $N_s/2$ offspring, denotes as $\{\mathcal{S}_{n_s} | n_s \in [N_s/2]\}$. After assessing the fitness score of $N_s/2$ offspring, we append them to $\Omega_n$.

\subsubsection{Long-Term Reflection (Step C)} \label{sec:4.2.4}
Upon the generation of $\{\theta_{n_s} | n_s \in [N_s/2]\}$, the long-term reflector LLM agent will aggregate them and ``previous long verbal gradient" $\Theta^{\prime}$ \footnote{This long verbal gradient is generated in the previous epoch.} into ``current long verbal gradient" $\Theta$. We present one instance of the ``long verbal gradient" as follows: ``\textit{Combine adaptive density-based clustering (e.g., DBSCAN) on inferred distributions with LP-enforced strict IR/IC constraints. Refine costs upward using rejection margins, normalize outcome probabilities, and prioritize feasibility. Employ silhouette scores and hierarchical methods to ensure robust, realistic agent settings matching all logs.}"

\subsubsection{Mutation (Step D)} \label{sec:4.2.5}
The mutation LLM agent leverages $\mathcal{S}^*$ and $\Theta$ as input to generate $N_m$ mutated P2 solvers, denoting as $\{\mathcal{S}_{n_m} | n_m \in [N_m]\}$. Subsequent to assessing the fitness score of mutations, we append them to $\Omega_n$. Moreover, we need to update the elitist P2 solver $\mathcal{S}^*$, revise the old population $\Omega_o = \Omega_n$, and clean the new population $\Omega_n$ before moving to the next epoch.

\subsection{Algorithm Design and Analysis} \label{sec:4.3}
Upon the preceding illustrations in Sections \ref{sec:4.1} and \ref{sec:4.2}, we present the LLM-empowered P2 solver evolution in Algorithm \ref{alg:2} for clarity. It is worth noting that the number of interaction rounds required for deriving a near-optimal contract by Algorithm \ref{alg:2} is $K+I$. Here, $K$ is the number of historical interaction logs as input to the P2 solver. $I$ is the number of iterations of Algorithm \ref{alg:2}. In each iteration, we need to evaluate the P2 solver via interaction between the teleoperator and the edge ASP.

\section{Experimental Evaluation} \label{sec:5}
In this section, we evaluate our proposed LLM-empowered online learning contract for bonus design. Specifically, we introduce the experimental configurations in Section \ref{sec:5.1}, analyze the scalability, sensitivity, and effectiveness of the LLM-empowered solution in Sections \ref{sec:5.2}, \ref{sec:5.3}, and \ref{sec:5.4}, respectively.

\subsection{Experimental Configurations} \label{sec:5.1}
\subsubsection{Parameters Configurations}
In this paper, we model the bonus design between the teleoperator and edge ASPs as an online learning contract design problem, described by elements $\mathcal{C} = (\mathcal{D}, \mathcal{R}, \mathcal{O}, \mathcal{A}, \Phi, q)$. We consider the size of historical interaction logs $|\mathcal{D}|=K=\{25, 50, 100\}$, size of outcome space $M=\{2, 4, 6, 8, 10, 12\}$, and the size of edge ASP's action space $N=\{2,3,4,5,6,7\}$. The outcome $o_m$ is an interval in $[0.9, 1.8]$, $\{o_m | m \in [2]\}=\{[0.9,1.35), [1.35,1.8)\}$ if $M=2$. The edge ASP's action $a_n$ is the diffusion step of the AIGC model, we set $a_n \in \{4,5,6,7,8,9,10\}$. The AIGC model is a conditional diffusion model, and we obtain it by referring to \cite{zhan2024vision}. Regarding the edge ASP's setting $\Phi=(\mathbf{P}, \mathbf{c})$, we utilize the statistical distribution on 200 AIGC tasks \cite{zhan2024vision, zhan2025distributionally} under the diffusion step $a_n$ as $\mathbf{p}_n$. Moreover, we consider the additional computational energy cost as $c_n$ under the diffusion step $a_n$, $\mathbf{c}= \$ \{0, 1.5^{-6}, 4^{-6}, 5.6^{-6}, 5.1^{-6}, 8.1^{-6}, 7.8^{-6}\}$ per image. Notably, we acquire the $\mathbf{c}$ via the electricity data from the Energy Information Administration (EIA) \footnote{https://www.eia.gov/electricity/monthly/update/end-use.php} and CodeCarbon \footnote{https://codecarbon.io/}. We set the AIGC services subscription fee per image as $r_s = \$ 1.6^{-4}$ by referring to the image pricing policy of Anthropic \footnote{https://docs.anthropic.com/en/docs/build-with-claude/vision}. With the assumption that the net profit of the AIGC model is around $30\%$, we set the AIGC model training cost spread per image to $ c_t=\$1.2^ {-4}$. We set the coefficient of mapping AIGC services quality to teleoperator's utility $\alpha \in \{5^{-4}, 1^{-3}, 5^{-3}\}$. All LLM agents in this paper are instances of gpt-4.1-mini-2025-04-14, and we set the iteration rounds $I=200$ by default.

\begin{figure*}[!t]
  \centering
  \begin{subfigure}[t]{0.3\textwidth}
    \centering
    \includegraphics[width=\linewidth]{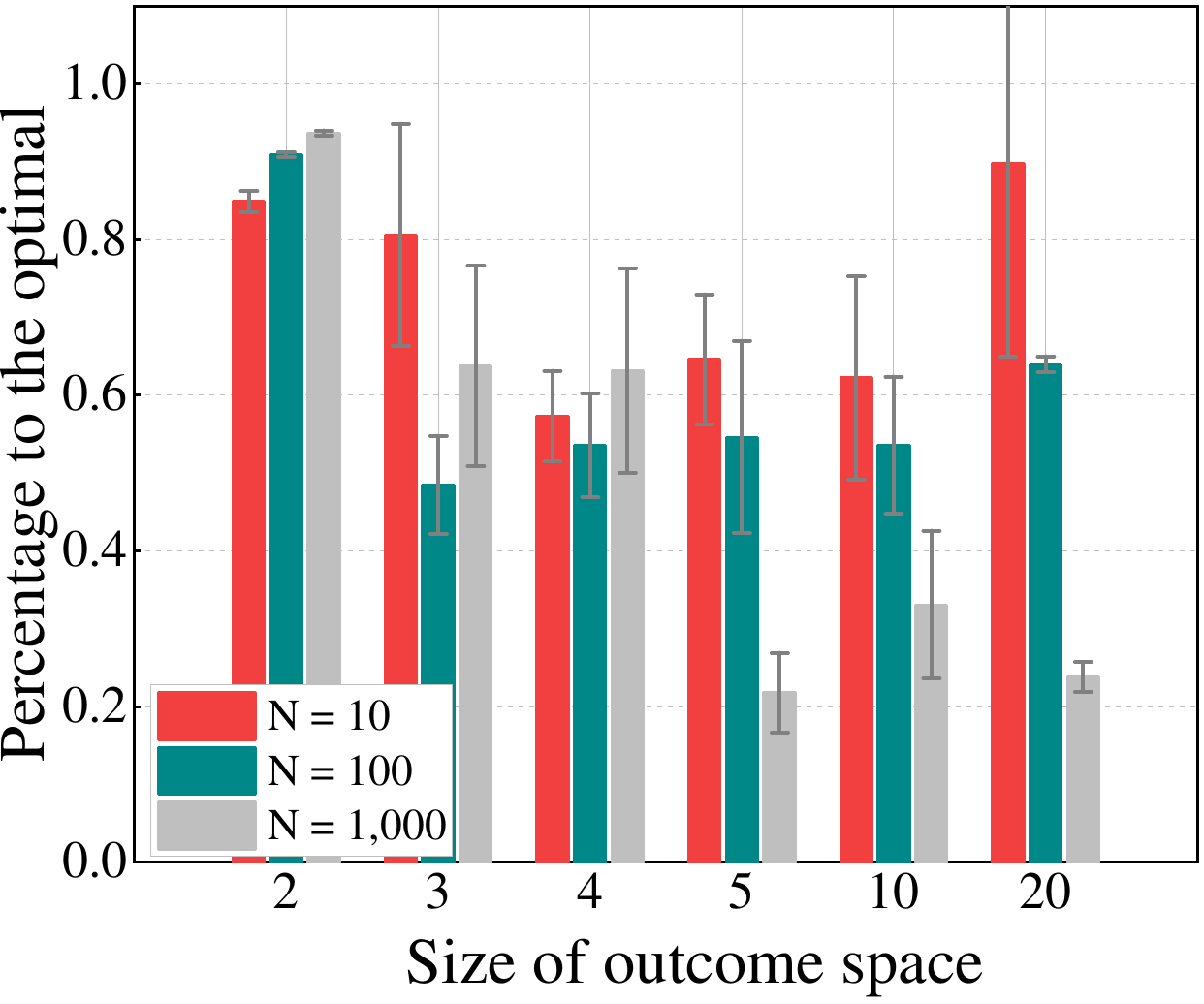}
    \caption{}
    \label{fig:4a}
  \end{subfigure}
  \hfill
  \begin{subfigure}[t]{0.3\textwidth}
    \centering
    \includegraphics[width=\linewidth]{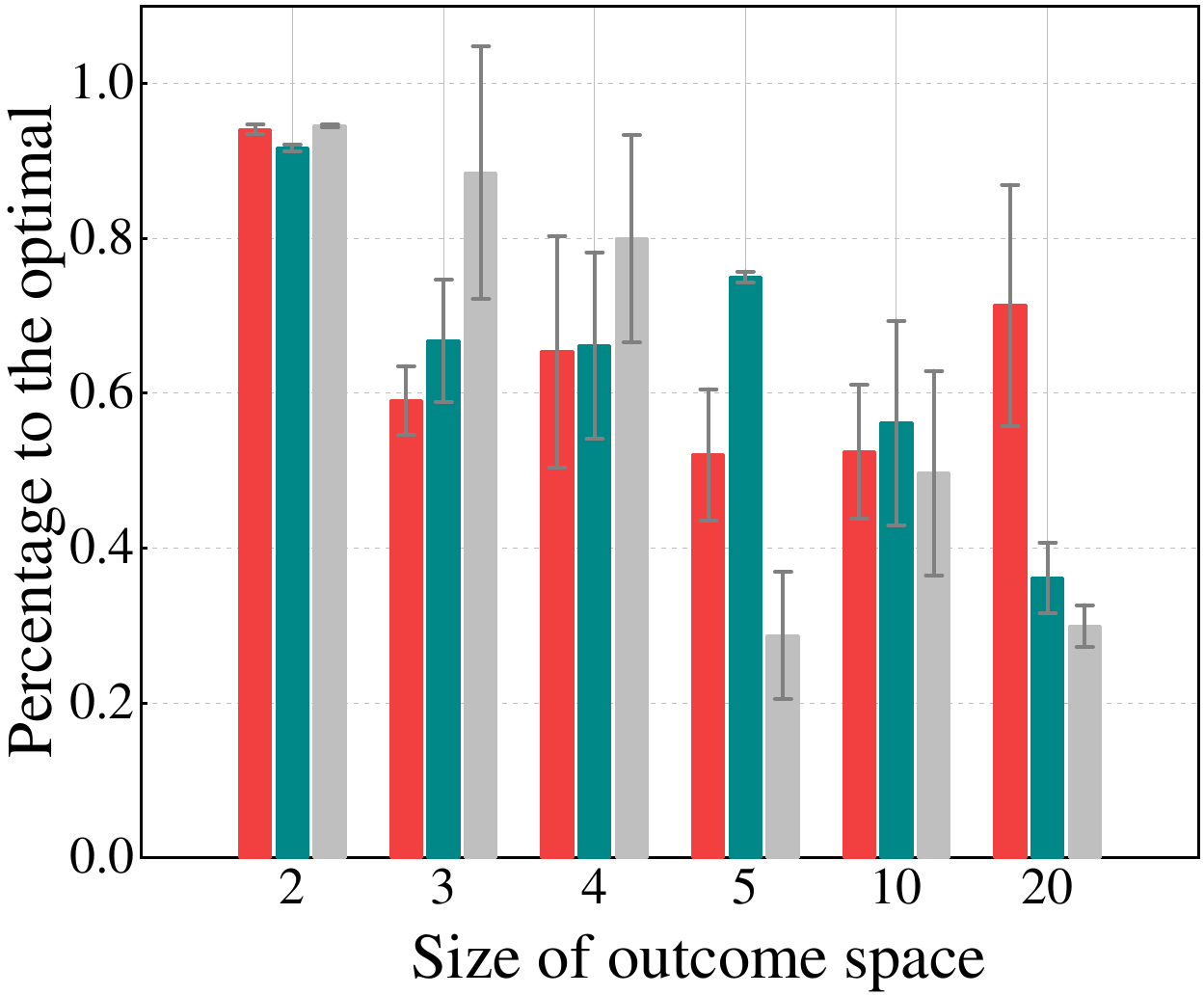}
    \caption{}
    \label{fig:4b}
  \end{subfigure}
  \hfill
  \begin{subfigure}[t]{0.3\textwidth}
    \centering
    \includegraphics[width=\linewidth]{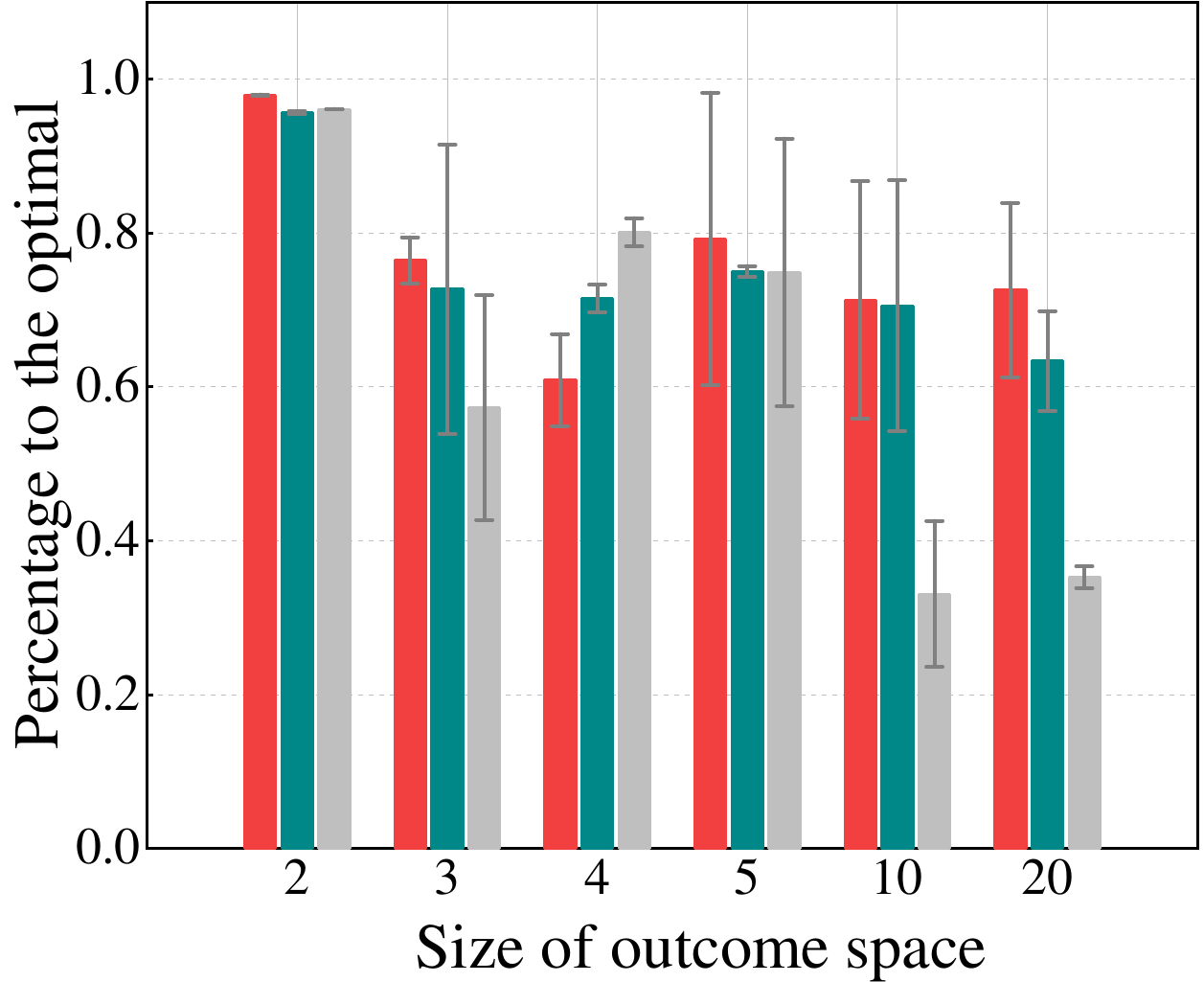}
    \caption{}
    \label{fig:4c}
  \end{subfigure}
  \caption{Evaluate the scalability of the proposed method by varying the size of the outcome space $M$, the number of actions $N$, and the number of historical interaction logs $K$, in which $K = \{25, 50, 100\}$ in Figs. \ref{fig:4a}, \ref{fig:4b}, and \ref{fig:4c}.}
  \label{fig:4}
\end{figure*}

\begin{figure*}[!t]
    \centering
    \begin{minipage}[t]{0.3\textwidth}
        \centering
        \includegraphics[width=\linewidth]{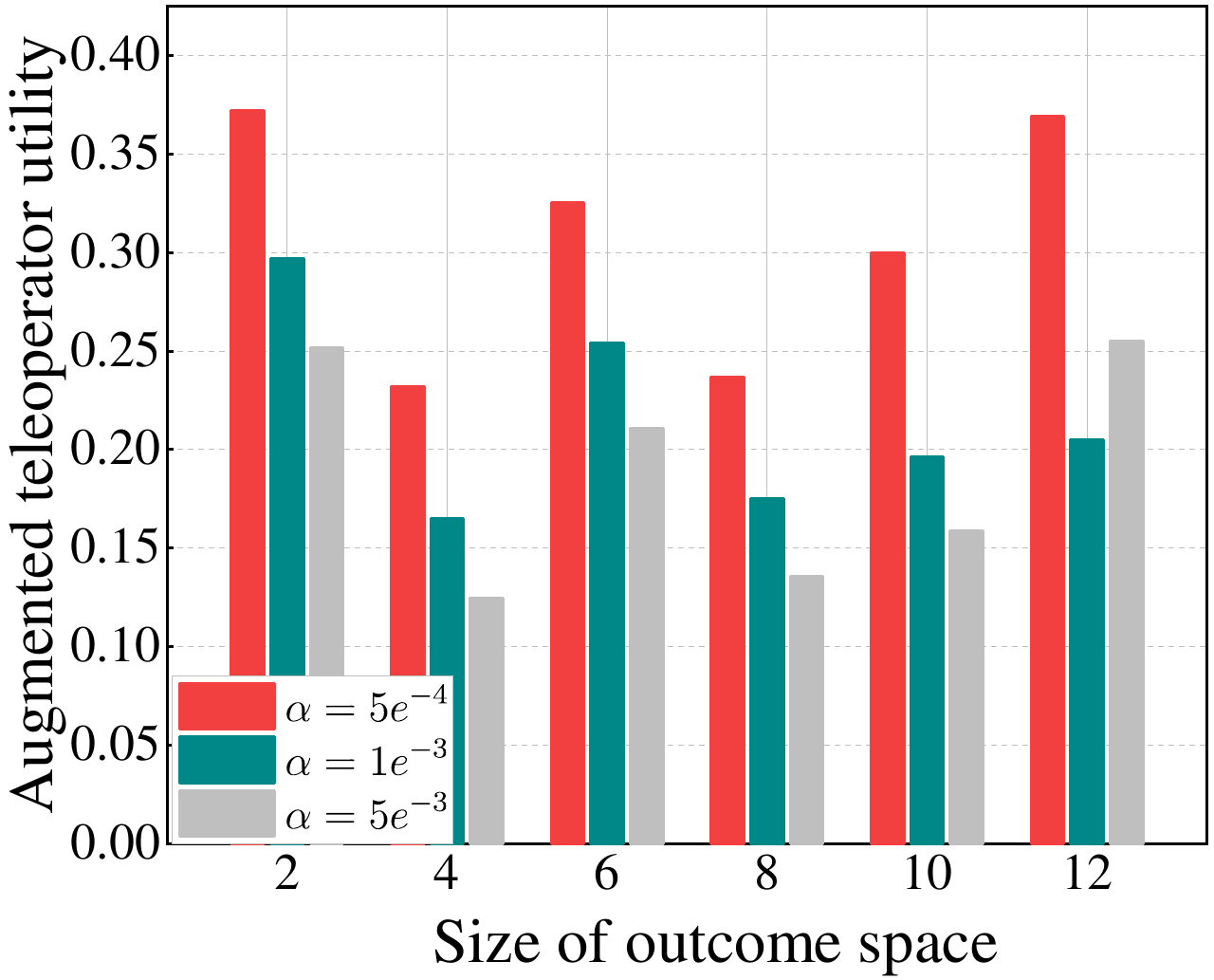}
        \caption{Assess the sensitivity of the proposed method regarding the mapping coefficient $\alpha$ in terms of $\pi_T^{\%}$.}
        \label{fig5}
    \end{minipage}
    \hfill
    \begin{minipage}[t]{0.3\textwidth}
        \centering
        \includegraphics[width=\linewidth]{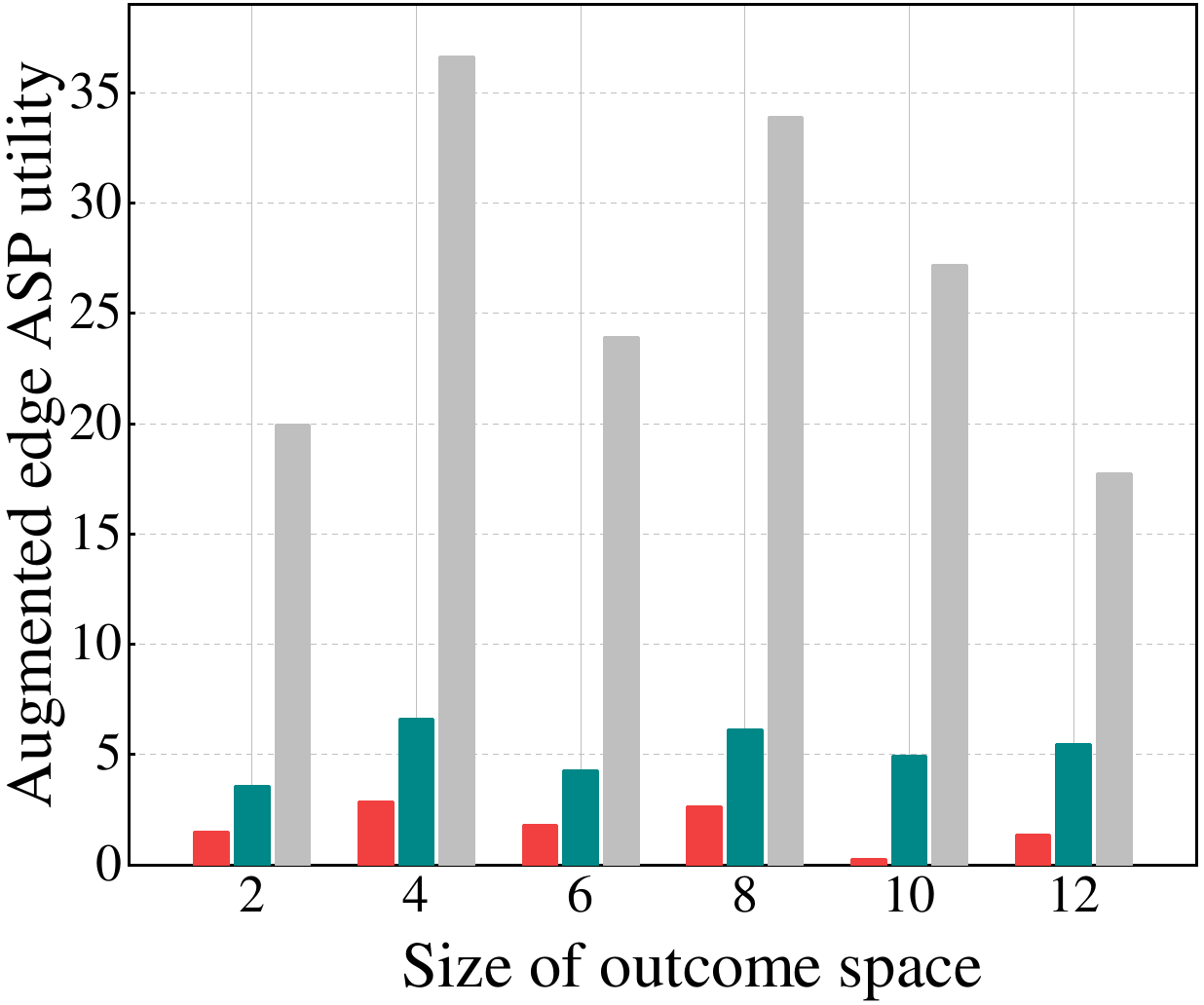}
        \caption{Assess the sensitivity of the proposed method regarding the mapping coefficient $\alpha$ in terms of $\pi_A^{\%}$.}
        \label{fig6}
    \end{minipage}
    \hfill
    \begin{minipage}[t]{0.3\textwidth}
        \centering
        \includegraphics[width=\linewidth]{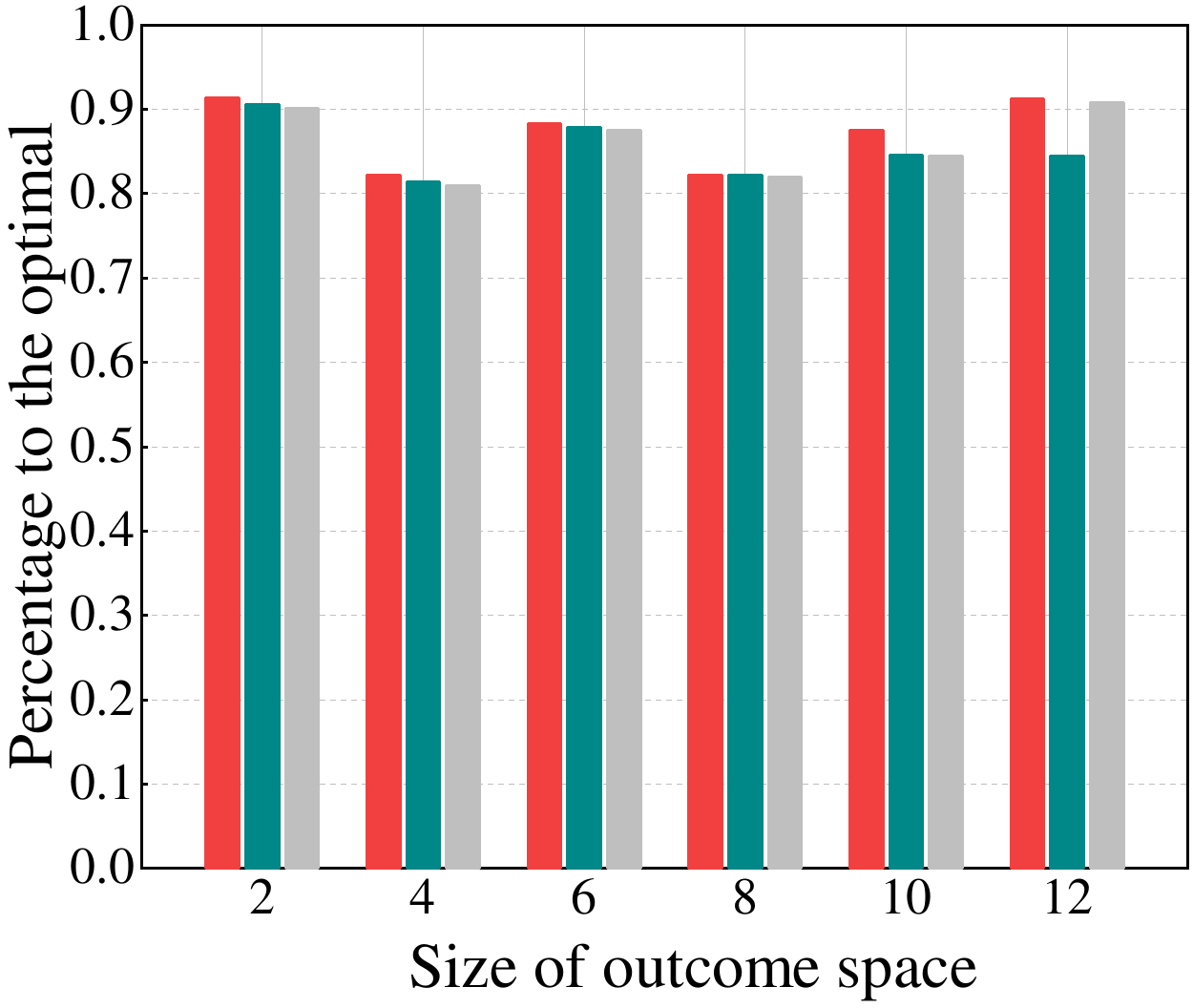}
        \caption{Assess the sensitivity of the proposed method regarding the mapping coefficient $\alpha$ in terms of $\eta$.}
        \label{fig7}
    \end{minipage}
\end{figure*}

\subsubsection{Benchmarks and Evaluation Metrics}
To assess the effectiveness of our proposed LLM-empowered online learning contract design, we compare it with the following benchmarks:
\begin{enumerate}
    \item [1)] \textbf{Seed}: We leverage the seed solver, i.e., Algorithm \ref{alg:1} to infer the edge ASP's setting $\tilde{\Phi}$, and then derive the contract via tackling P3. With this benchmark, we can evaluate the effectiveness of our proposed solution.
    \item [2)] \textbf{Zero-shot}: We run the Algorithm \ref{alg:2} under a simulated scenario, and utilize the output $\hat{\mathcal{S}}^*$ as the P2 solver for teleoperator. In this way, we can assess the performance of our proposed solution in a limited $K$ interaction rounds between the teleoperator and the edge ASP.
    \item [3)] \textbf{Bandit}: By referring to \cite{ho2014adaptive, zhu2022sample}, we use the multi-armed bandit method to derive the contract directly.
\end{enumerate}

In this paper, we evaluate our proposed solution in three metrics, augmented teleoperator utility $\pi_{T}^{\%}$, augmented edge ASP utility $\pi_A^{\%}$, and percentages to the optimal $\eta$. Without a bonus incentive mechanism (contract), we consider the edge ASP will take $a_1 = 4$ diffusion steps by default. Therefore, we define $\pi_{T}^{\%}$ as:
\begin{equation} \label{eq:10}
    \pi_{T}^{\%} = (\pi_T(\tilde{\mathbf{r}}) - \pi_T(\mathbf{0})) / \pi_T(\mathbf{0}).
\end{equation}
Analogously, we define $\pi_A^{\%}$ as:
\begin{equation} \label{eq:11}
    \pi_A^{\%} = (\pi_A(a_i;\pi_T(\tilde{\mathbf{r}}),\Phi) - \pi_A(a_1;\mathbf{0},\Phi)) /  \pi_A(a_1;\mathbf{0},\Phi).
\end{equation}
Regarding the metric of percentages to the optimal $\eta$, we define it as:
\begin{equation} \label{eq:12}
    \eta = \pi_T(\tilde{\mathbf{r}}) / \pi_T(\mathbf{r}^*),
\end{equation}
where $\mathbf{r}^*$ is the optimal contract under the assumption that the teleoperator is privy to the edge ASP's setting $\Phi$.

We utilize Python to run all simulations in this paper and conduct the experiments on a laptop equipped with a 6-core AMD Ryzen 5 processor, 32 GB of RAM, an RTX 4060 GPU, and a Windows 11 operating system. We ran each experiment multiple times to enhance the reliability of the result and take the average value as the final result.

\subsection{Scalability of LLM-Empowered Solution} \label{sec:5.2}
In this section, we simulate an online learning contract problem, as described in \cite{wang2023deep}, to evaluate the scalability of our proposed method. Concretely, we apply the SoftMax on a Gaussian random vector in $\mathbb{R}^{M}$ to generate the outcome distributions $\mathbf{p}_n$ for action $a_n$. We sample the outcome $q_m$ uniformly from $[0, 10]$. We set the action cost as a mixture, $c_n = (1-\beta_p)c_r(a_n)+\beta_p c_i(a_n)$, where $c_r(a_n) = \beta_c \mathbb{E}_{o_m\sim\mathbf{p}_n} {[q_m]}$ is a correlated cost that proportional to the expected value of the action. $c_i(a_n)$ is an independent cost and uniform on $[0, 1]$. We set the coefficients $\beta_c=0.7$ and $\beta_p = 0.3$ by default. To assess the scalability, we run experiments under the size of outcome space $M \in \{2, 3, 4, 5, 10, 20\}$, the size of action space $N \in \{10, 100, 1000\}$, and the number of historical interaction logs $K \in \{25, 50, 100\}$.

In light of our use of the simulated dataset to assess the scalability of our proposed method, we focus solely on the metric $\eta$ and present the evaluation results in Fig. \ref{fig:4}. Observing Figs. \ref{fig:4a}, \ref{fig:4b}, and \ref{fig:4c} show that the performance of our proposed method increases with the number of interaction logs. Moreover, as $M$ and $N$ increase, the performance of our proposed method declines. Remarkably, observing Fig. \ref{fig:4c}, our proposed method can maintain $\eta=75\%$ under varying $M$ and $N$ when $K=100$. The performance of our proposed method can even maintain $\eta=95\%$ under varying sizes of action space when $M=2$ and $K=\{50, 100\}$. Therefore, our proposed method possesses scalability in terms of the size of the action space $N$ under a small number of interaction rounds ($50$) between the edge ASP and the teleoperator.

\begin{figure*}[!t]
  \centering
  \begin{subfigure}[t]{0.3\textwidth}
    \centering
    \includegraphics[width=\linewidth]{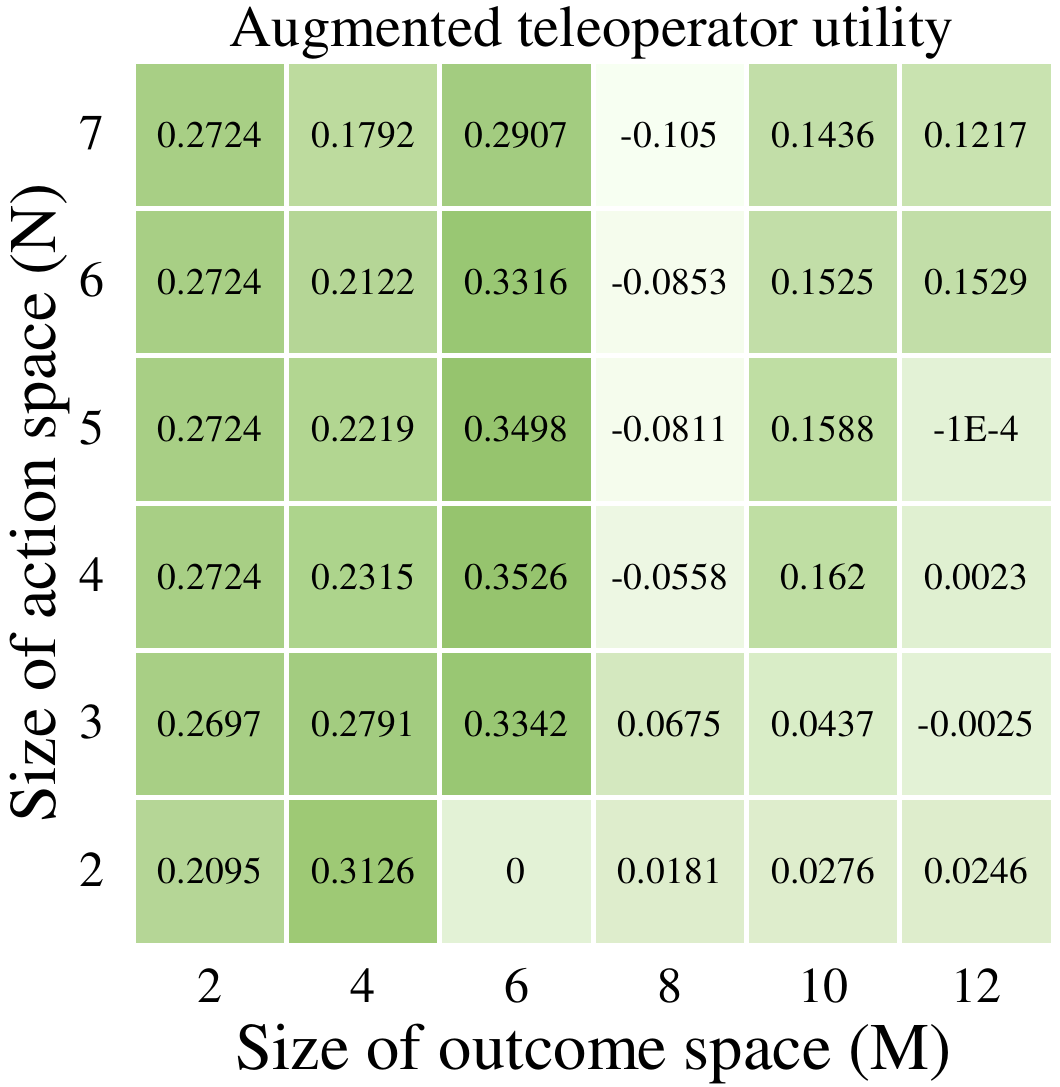}
    \caption{}
    \label{fig:8a}
  \end{subfigure}
  \hfill
  \begin{subfigure}[t]{0.3\textwidth}
    \centering
    \includegraphics[width=\linewidth]{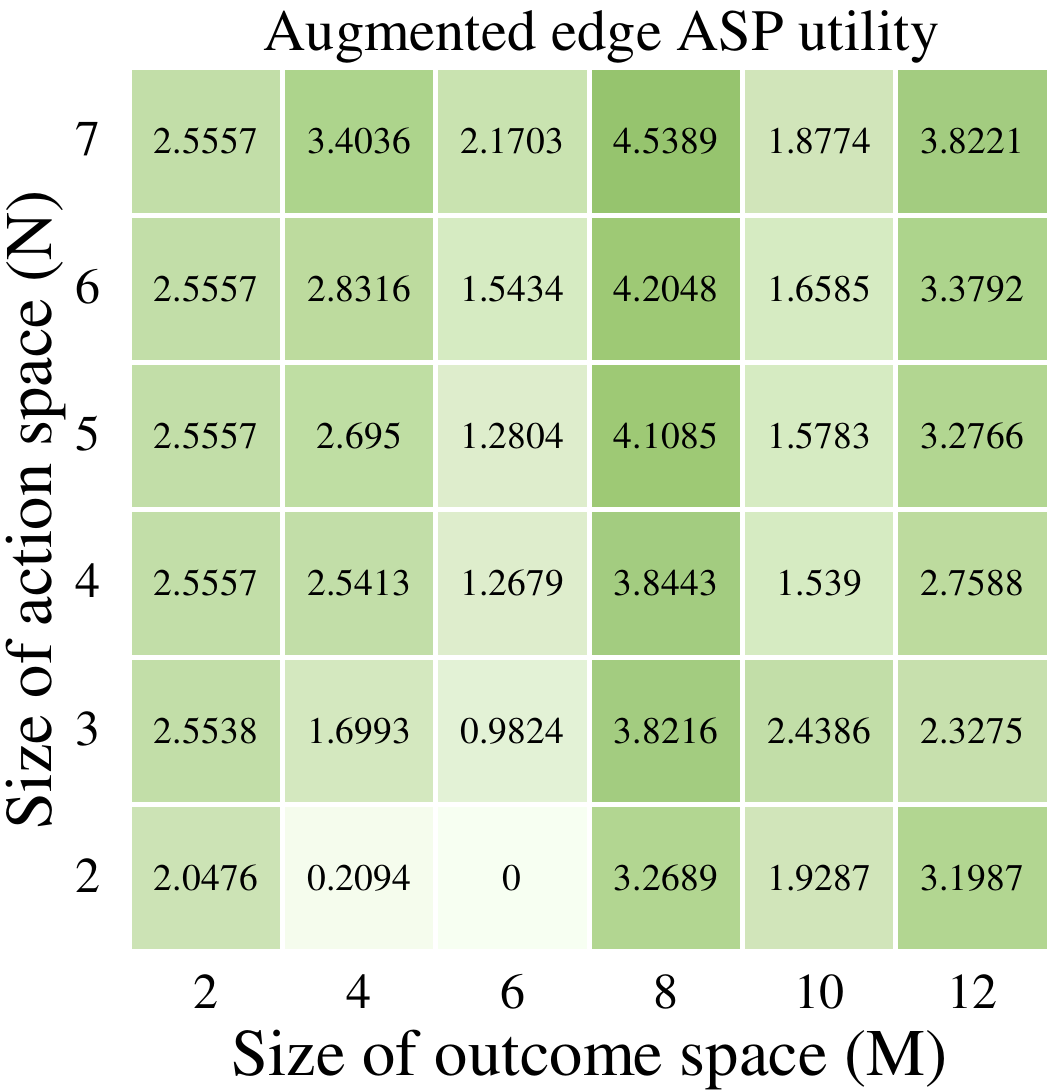}
    \caption{}
    \label{fig:8b}
  \end{subfigure}
  \hfill
  \begin{subfigure}[t]{0.3\textwidth}
    \centering
    \includegraphics[width=\linewidth]{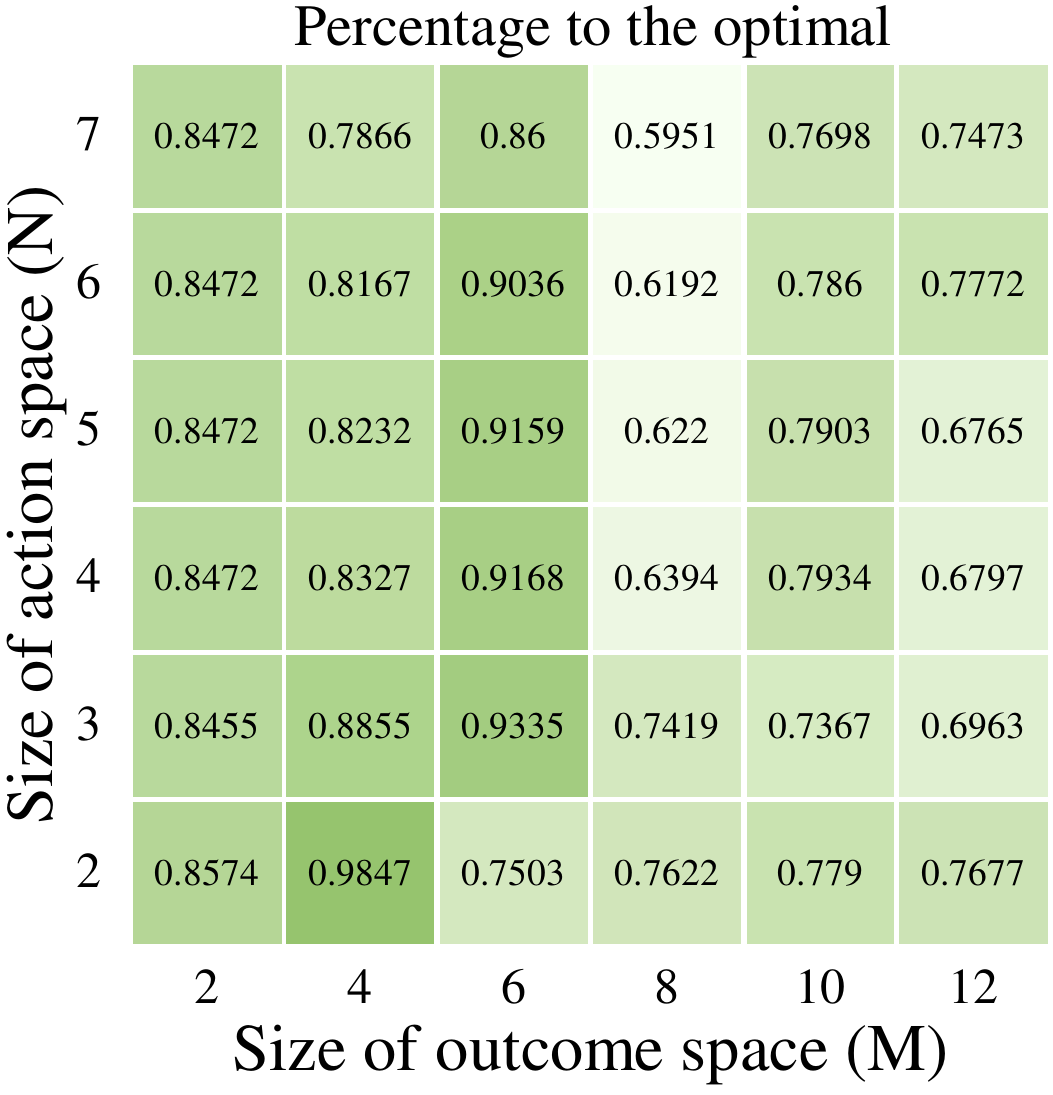}
    \caption{}
    \label{fig:8c}
  \end{subfigure}
  \caption{Assess the sensitivity of the proposed method regarding the size of outcome space $M$ and the size of action space $N$ in terms of $\pi_T^{\%}$, $\pi_A^{\%}$, and $\eta$. We set the number of historical interaction logs $K=25$.}
  \label{fig:8}
\end{figure*}

\begin{figure*}[!t]
  \centering

  \begin{subfigure}[t]{0.3\textwidth}
    \centering
    \includegraphics[width=\linewidth]{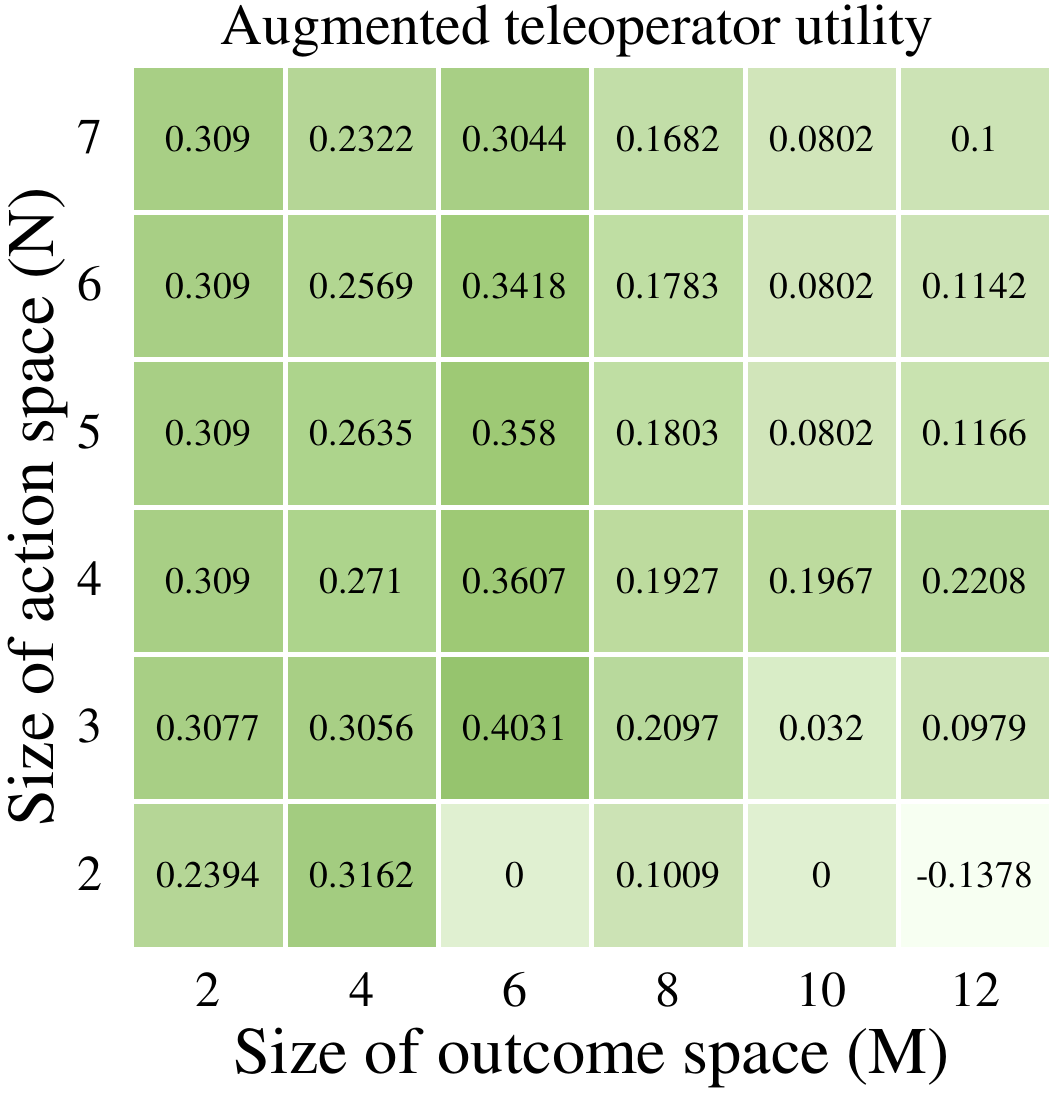}
    \caption{}
    \label{fig:9a}
  \end{subfigure}\hfill
  \begin{subfigure}[t]{0.3\textwidth}
    \centering
    \includegraphics[width=\linewidth]{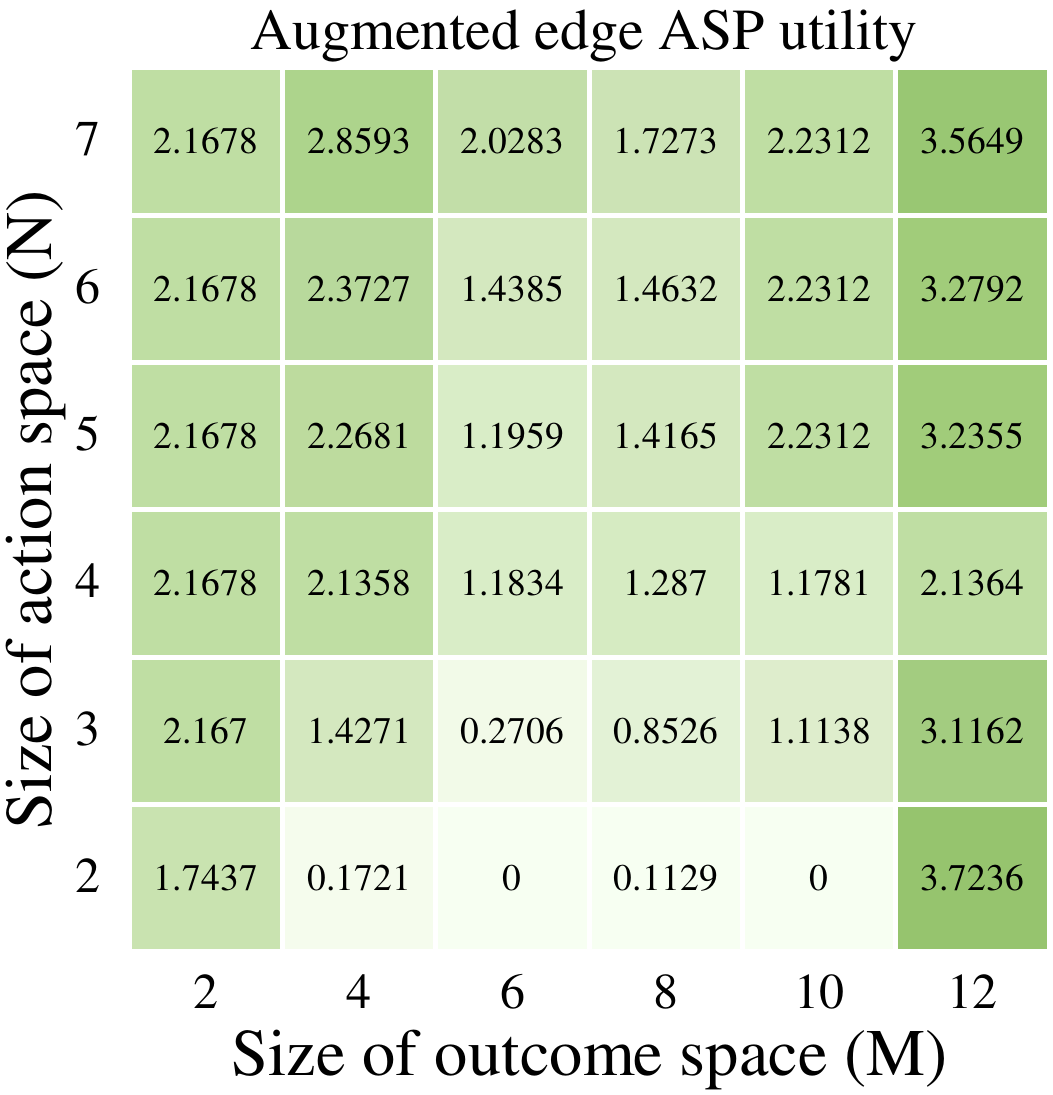}
    \caption{}
    \label{fig:9b}
  \end{subfigure}\hfill
  \begin{subfigure}[t]{0.3\textwidth}
    \centering
    \includegraphics[width=\linewidth]{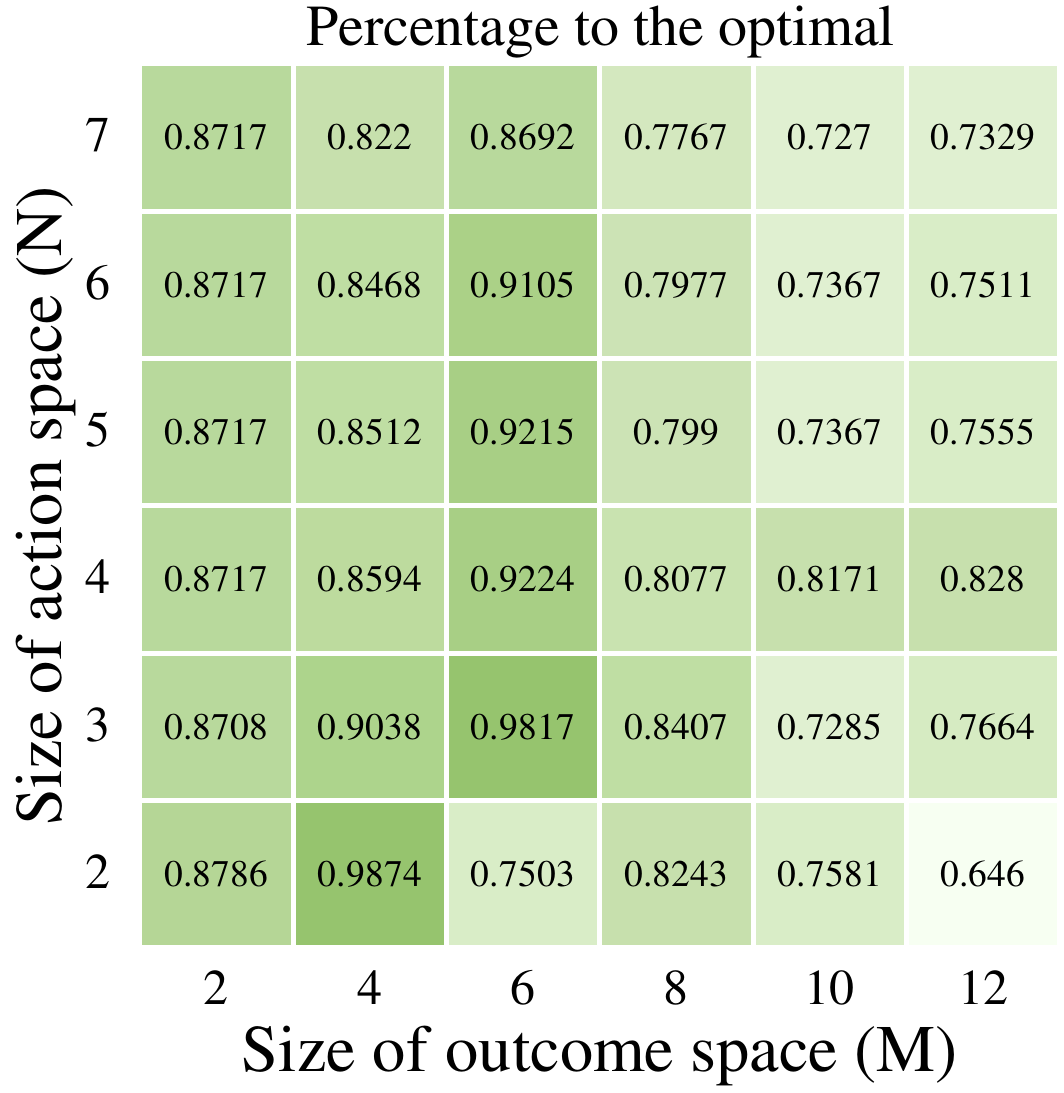}
    \caption{}
    \label{fig:9c}
  \end{subfigure}

  \caption{Assess the sensitivity of the proposed method regarding the size of outcome space $M$ and the size of action space $N$ in terms of $\pi_T^{\%}$, $\pi_A^{\%}$, and $\eta$. We set the number of historical interaction logs $K=50$.}
  \label{fig:9}
\end{figure*}

\subsection{Sensitivity of LLM-Empowered Solution} \label{sec:5.3}

In this section, we evaluate the sensitivity of our proposed method regarding the bonus design (contract) for edge ASP in terms of the mapping coefficient $\alpha$, the size of the outcome space $M$, the size of the action space $N$, and the number of historical interaction logs $K$. Remarkably, we set $\alpha=5^{-4}$, $M = 12$, $N = 7$, and $K=100$ by default if without other specifications.

As shown in Figs. \ref{fig5}, \ref{fig6}, and \ref{fig7}, we present the experimental results regarding $\pi_T^{\%}$, $\pi_A^{\%}$, and $\eta$ by varying the mapping coefficient $\alpha$. Observing Fig. \ref{fig5}, $\pi_T^{\%}$ is diminishing along with the increasing of $\alpha$. The basic rationale is that the teleoperator places higher importance on AIGC service quality and is willing to offer a higher bonus, thereby eliciting the edge that ASP takes more diffusion steps. The results in Fig. \ref{fig6} demonstrate this point, where the edge ASP's augmented utility is escalating along with the increase in $\alpha$. By analyzing Fig. \ref{fig7}, our proposed method is robust against $\alpha$, which maintains $\eta=90\%$ under varying $\alpha$.

\begin{figure*}[!t]
  \centering

  \begin{subfigure}[t]{0.3\textwidth}
    \centering
    \includegraphics[width=\linewidth]{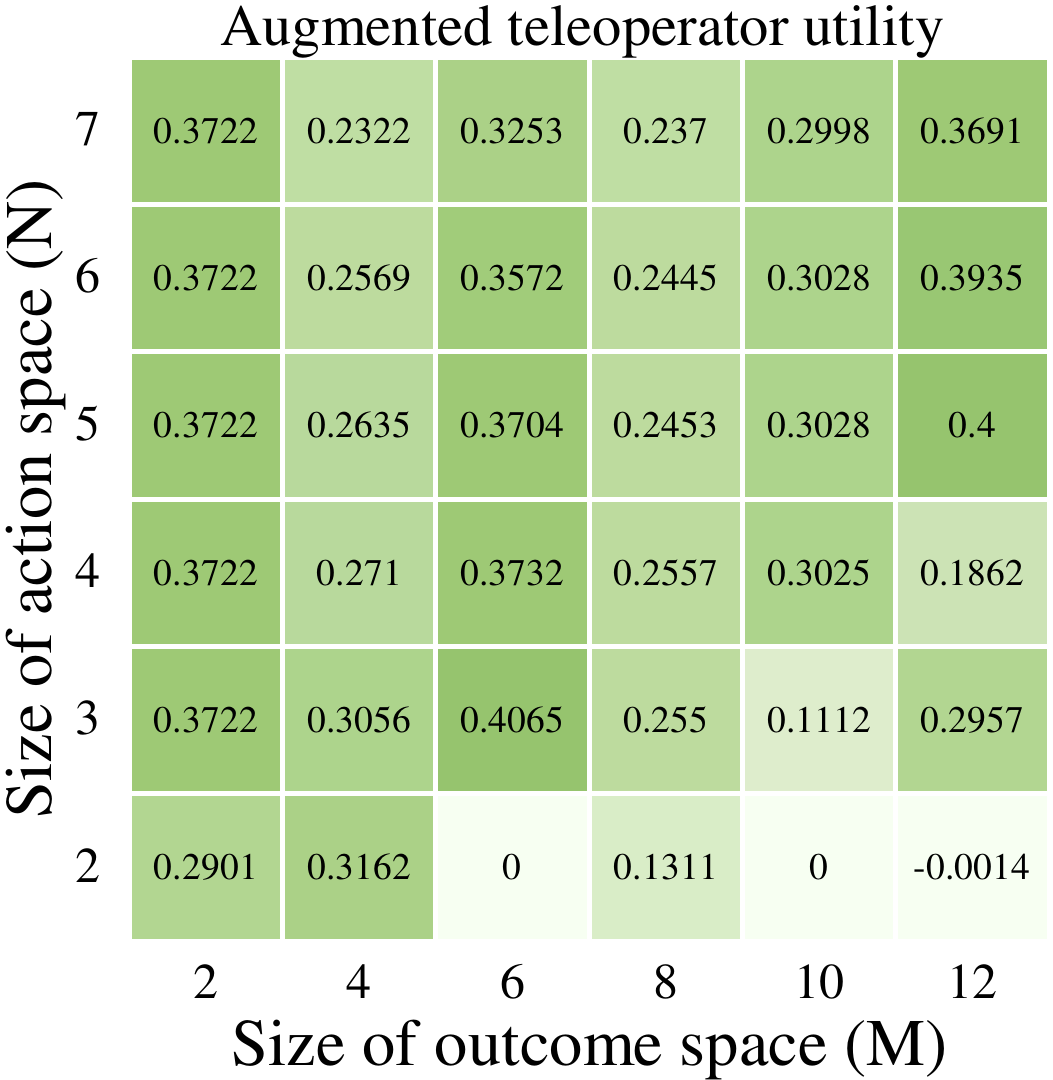}
    \caption{}
    \label{fig:10a}
  \end{subfigure}\hfill
  \begin{subfigure}[t]{0.3\textwidth}
    \centering
    \includegraphics[width=\linewidth]{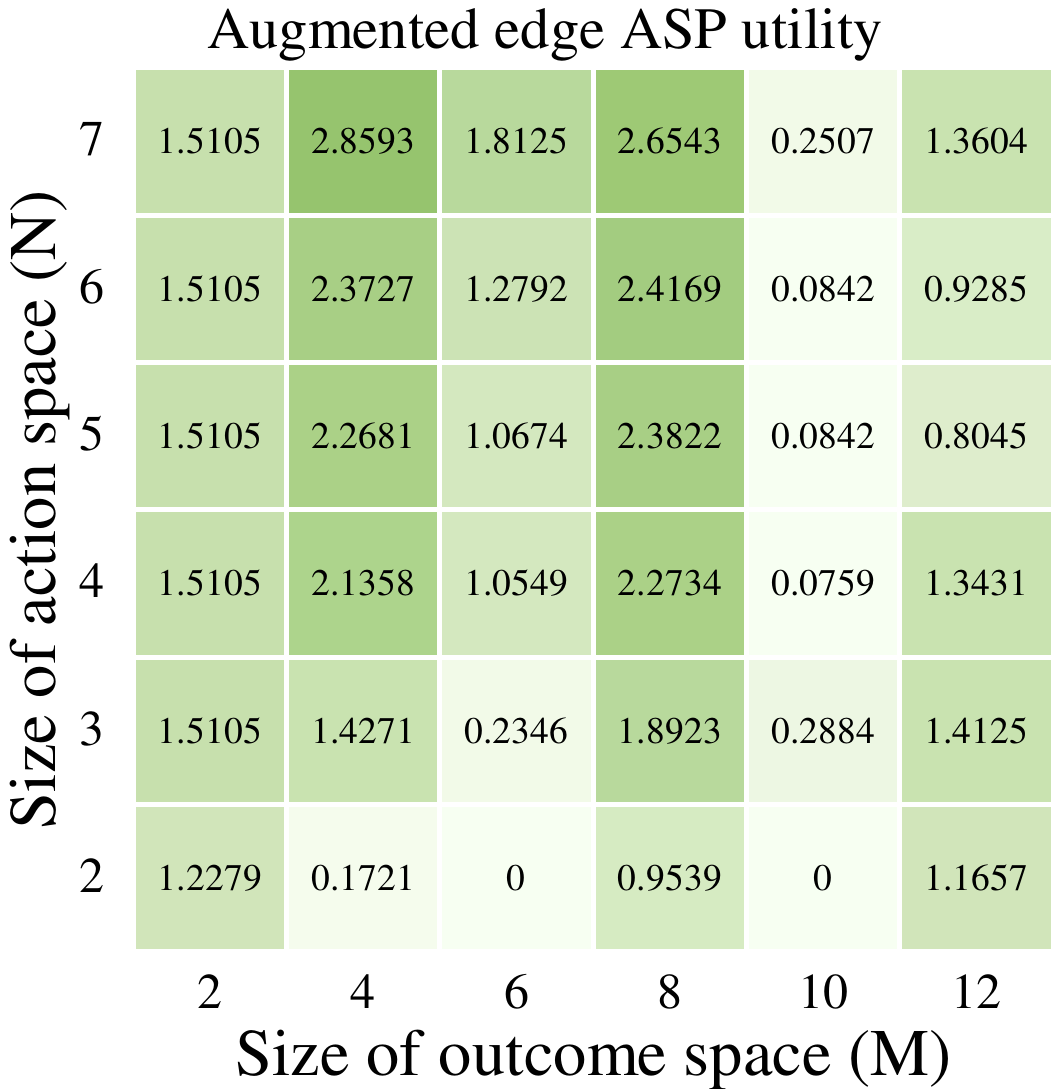}
    \caption{}
    \label{fig:10b}
  \end{subfigure}\hfill
  \begin{subfigure}[t]{0.3\textwidth}
    \centering
    \includegraphics[width=\linewidth]{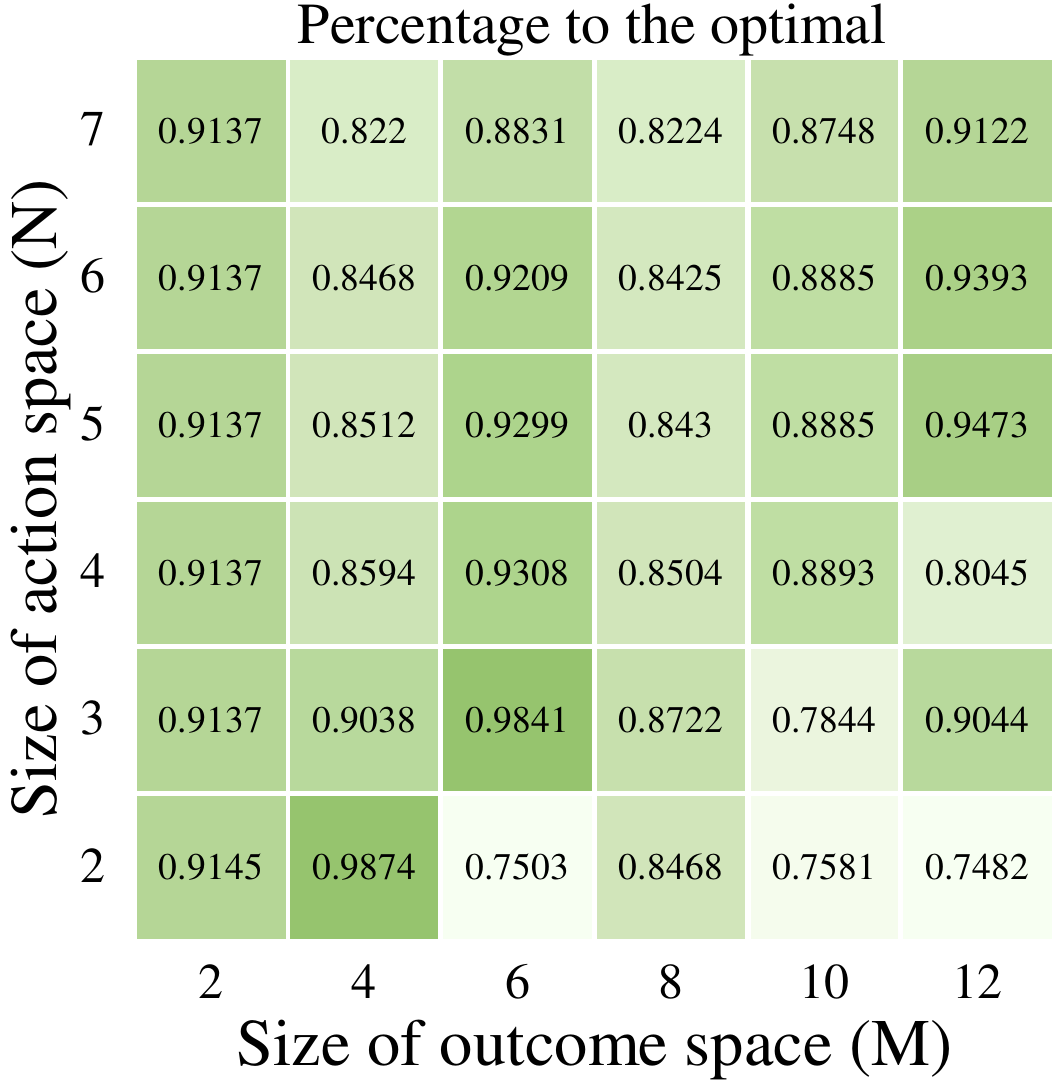}
    \caption{}
    \label{fig:10c}
  \end{subfigure}

  \caption{Assess the sensitivity of the proposed method regarding the size of outcome space $M$ and the size of action space $N$ in terms of $\pi_T^{\%}$, $\pi_A^{\%}$, and $\eta$. We set the number of historical interaction logs $K=100$.}
  \label{fig:10}
\end{figure*}

In Fig. \ref{fig:8}, we set $K=25$ and vary $M$ and $N$ simultaneously to analyze the sensitivity of our proposed method. As shown in Fig. \ref{fig:8a}, the mounting of $M$ will deteriorate $\pi_T^{\%}$, which means our proposed method is sensitive to $M$. However, $\pi_T^{\%}$ is escalating along with the increase in $N$; our proposed method is robust against $N$. The basic rationale is that the rise of $M$ will cause the size of edge ASPs' setting $\Phi$ to increase, thereby improving the difficulty in deriving a near-optimal contract. Notably, although the escalations of $N$ will also increase the size of $\Phi$, the mounted action space indicates that more fine-grained actions are available for the contract offered by the teleoperator. The results in Figs. \ref{fig:8a} and $\ref{fig:8b}$ are akin to the analysis of Figs. \ref{fig5} and \ref{fig6}, i.e., increasing $\pi_A^{\%}$ deteriorates $\pi_T^{\%}$ since more utility is transferred from the teleoperator to the edge ASP via contract. The results in Figs. \ref{fig:9} and \ref{fig:10} are akin to Fig. \ref{fig:8}. Notably, the performance of our proposed method is increasing along with the escalation of $K$, and our proposed method can achieve around $\eta=90\%$ when $K=100$ under varying $M$ and $N$ as the results depicted in Fig. \ref{fig:10c}.

\subsection{Effectiveness of LLM-Empowered Solution} \label{sec:5.4}

In this section, we assess the effectiveness of our proposed method by comparing it with benchmarks, varying $\alpha$, $M$, and $N$. Notably, the benchmarks in this experiment comprise $K=300$ historical interaction logs for contract derivation, as our proposed method utilizes $K=100$ historical interaction logs and $I=200$ iteration rounds for Algorithm \ref{alg:2}.

\begin{figure*}[!t]
    \centering
    \begin{minipage}[t]{0.3\textwidth}
        \centering
        \includegraphics[width=\linewidth]{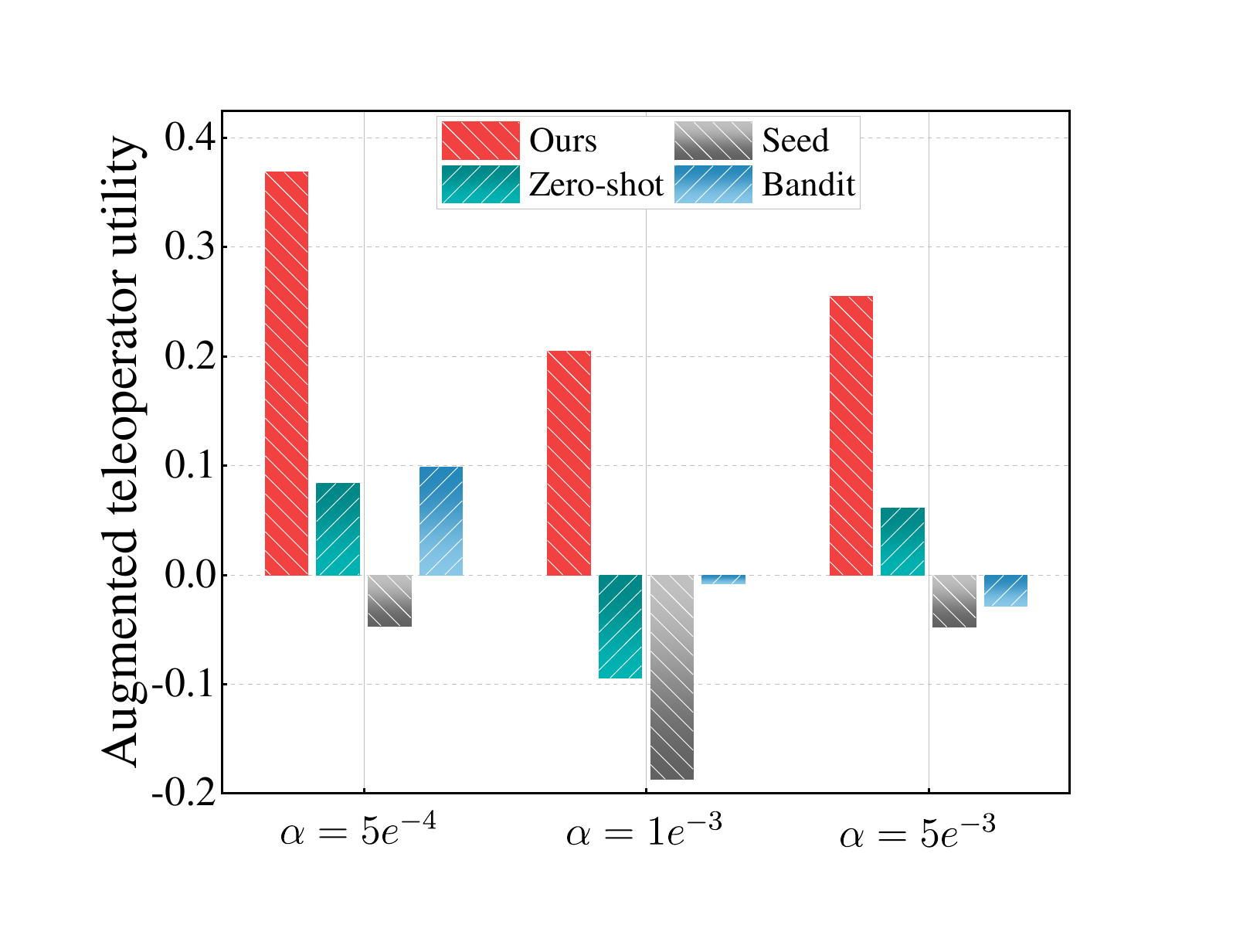}
        \caption{Comparison results by varying the mapping coefficient $\alpha$ in terms of $\pi_T^{\%}$.}
        \label{fig11}
    \end{minipage}
    \hfill
    \begin{minipage}[t]{0.3\textwidth}
        \centering
        \includegraphics[width=\linewidth]{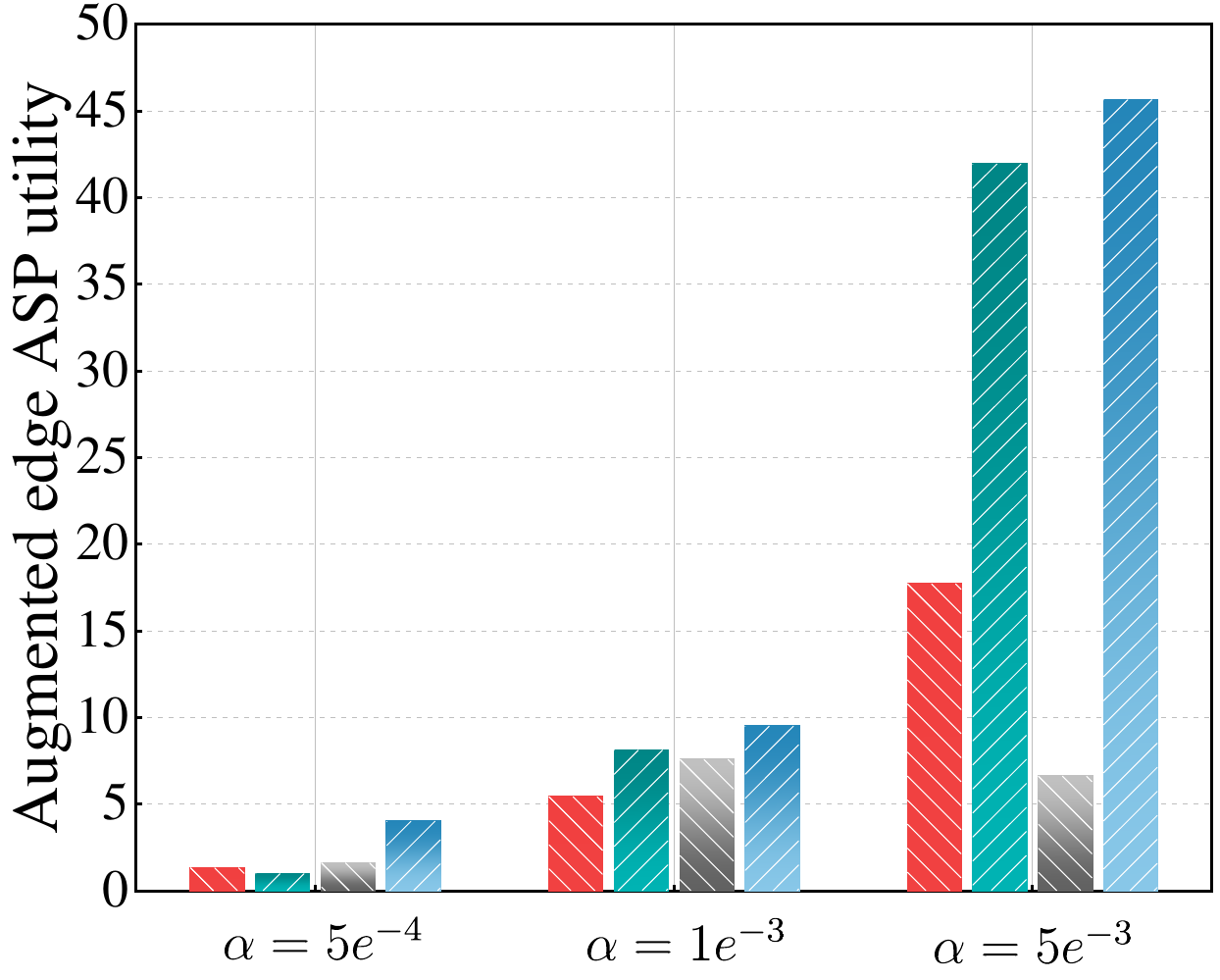}
        \caption{Comparison results by varying the mapping coefficient $\alpha$ in terms of $\pi_A^{\%}$.}
        \label{fig12}
    \end{minipage}
    \hfill
    \begin{minipage}[t]{0.3\textwidth}
        \centering
        \includegraphics[width=\linewidth]{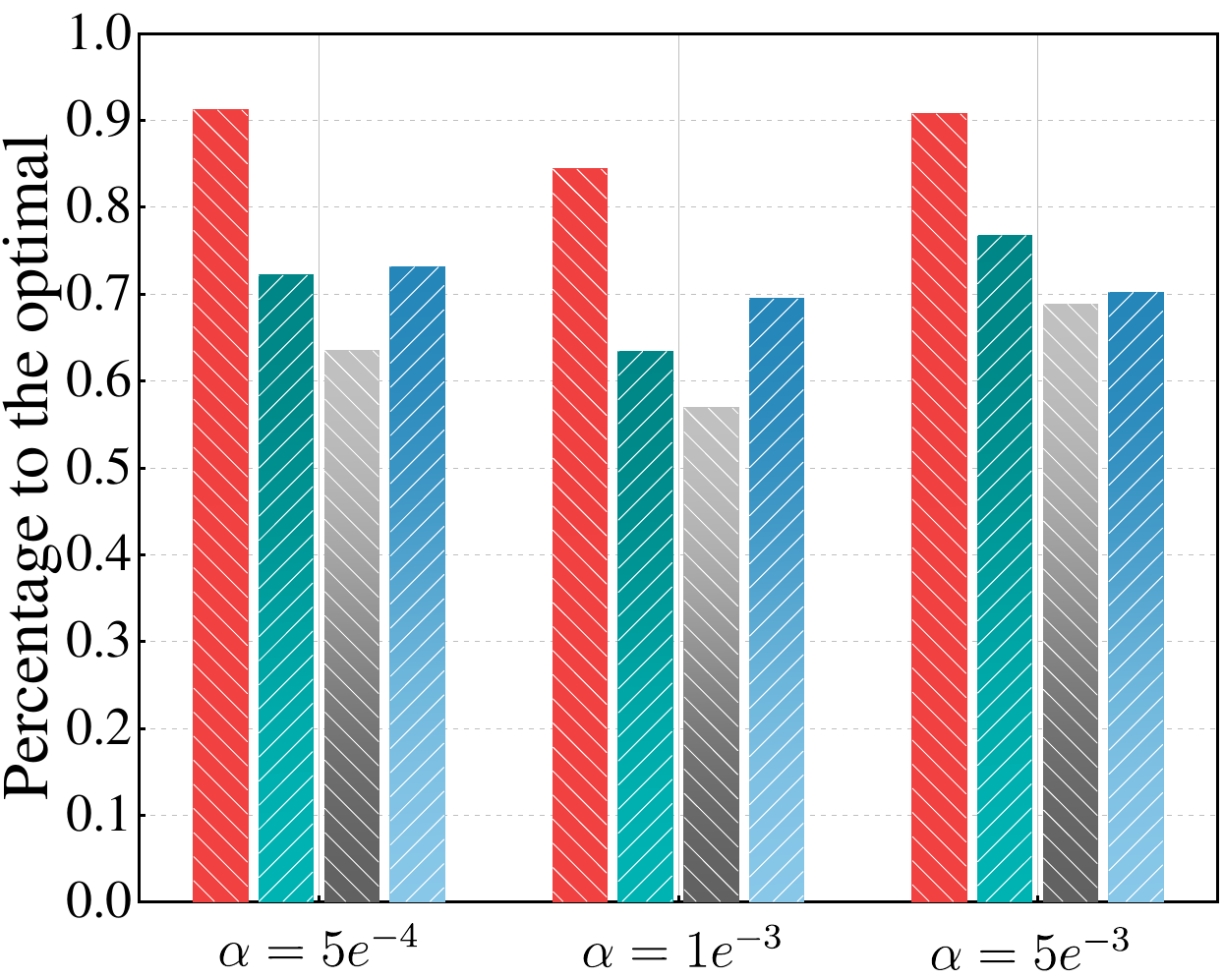}
        \caption{Comparison results by varying the mapping coefficient $\alpha$ in terms of $\eta$.}
        \label{fig13}
    \end{minipage}
\end{figure*}

Observing Fig. \ref{fig11}, our proposed method consistently surpasses the benchmarks in terms of $\pi_T^{\%}$ under varying $\alpha$, which demonstrates the effectiveness of our proposed method. Compared to the seed solver, our proposed method can augment $\pi_T^{\%}$ by around $40\%$, further demonstrating the effectiveness of our proposed method. Compared with the traditional bandit algorithm, our proposed method can augment $\pi_T^{\%}$ around $20\sim25\%$. As depicted in Fig. \ref{fig12}, although our proposed method yields the worst improvement in edge ASP utility, it can still ensure the IR constraint, thereby incentivizing the edge ASP to supply high-quality AIGC services at a minimal cost. The results in Fig. \ref{fig13} are similar to those in Fig. \ref{fig11}; our proposed method outperforms the benchmarks and maintains $\eta$ at around $90\%$.

\begin{figure*}[!t]
    \centering
    \begin{minipage}[t]{0.3\textwidth}
        \centering
        \includegraphics[width=\linewidth]{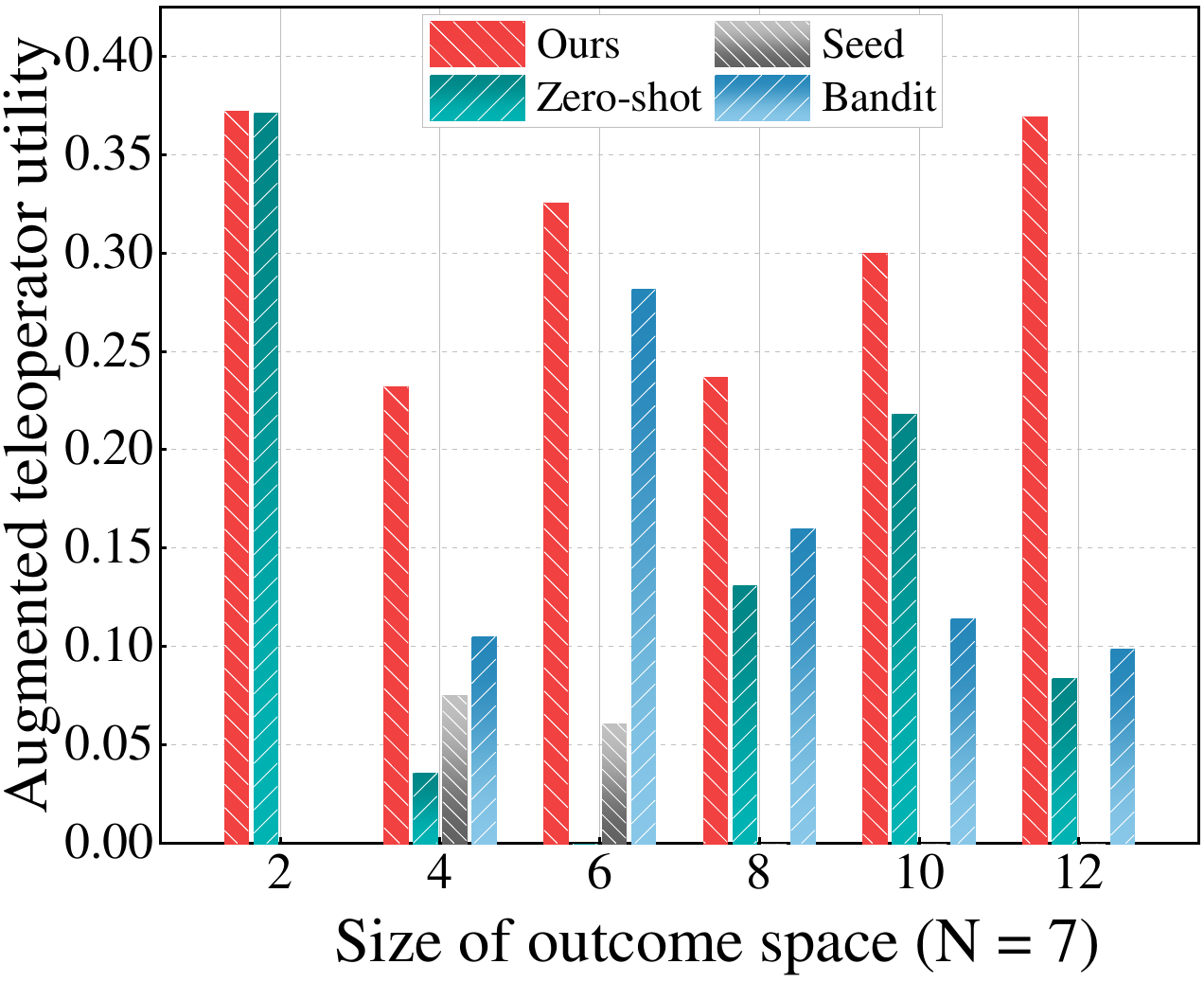}
        \caption{Comparison results by varying the size of the outcome space $M$ in terms of $\pi_T^{\%}$.}
        \label{fig14}
    \end{minipage}
    \hfill
    \begin{minipage}[t]{0.3\textwidth}
        \centering
        \includegraphics[width=\linewidth]{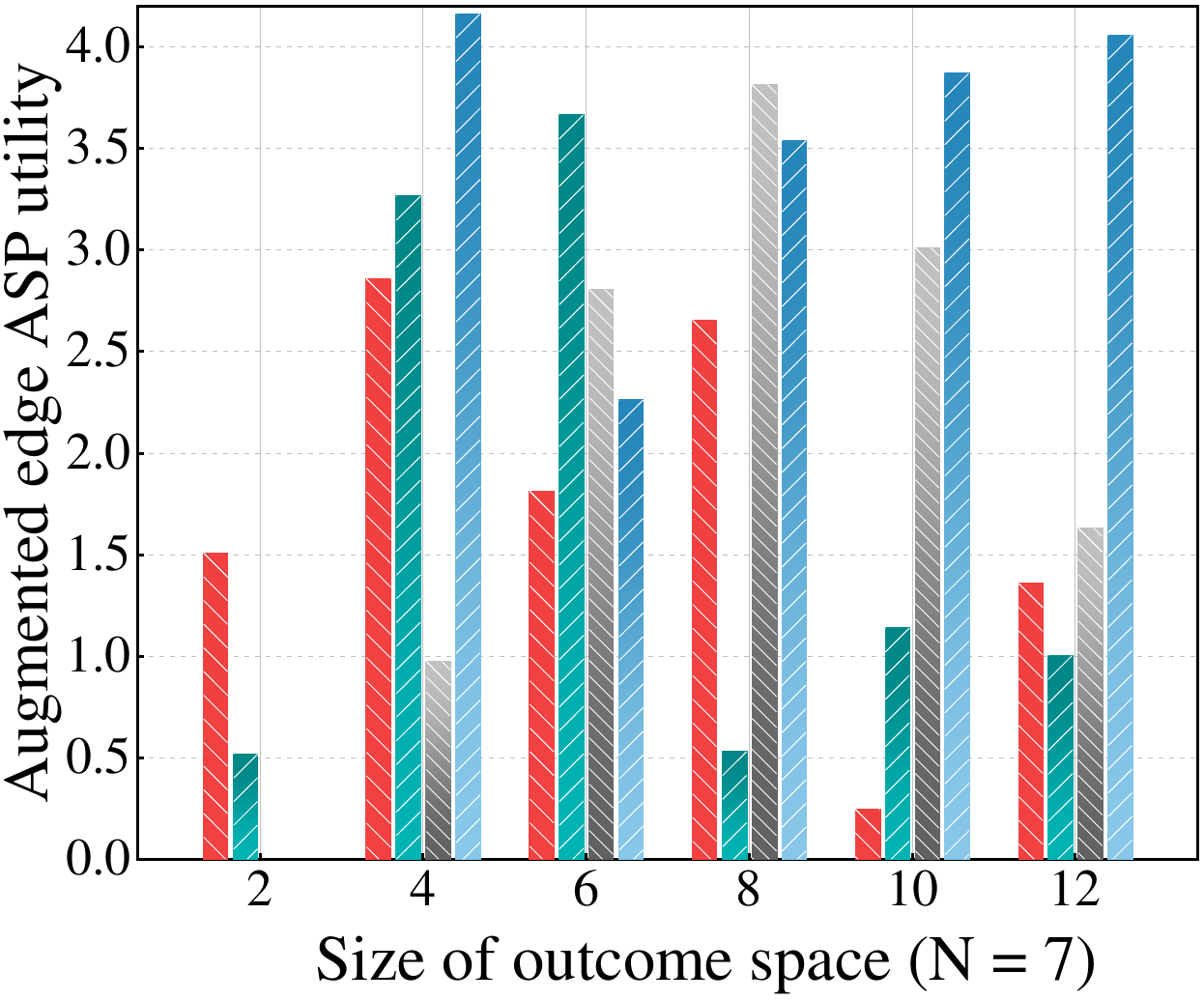}
        \caption{Comparison results by varying the size of the outcome space $M$ in terms of $\pi_A^{\%}$.}
        \label{fig15}
    \end{minipage}
    \hfill
    \begin{minipage}[t]{0.3\textwidth}
        \centering
        \includegraphics[width=\linewidth]{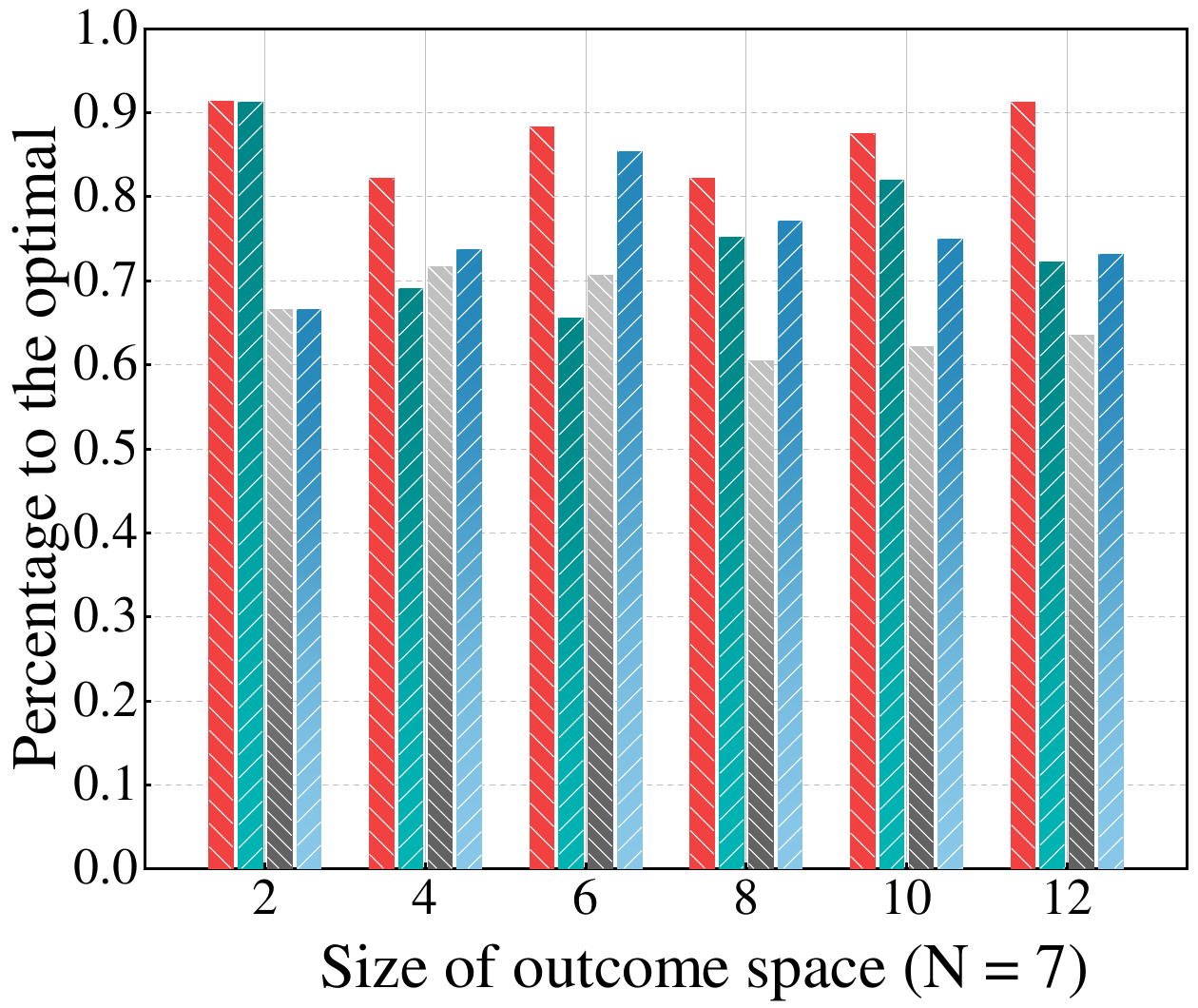}
        \caption{Comparison results by varying the size of the outcome space $M$ in terms of $\eta$.}
        \label{fig16}
    \end{minipage}
\end{figure*}

\begin{figure*}[!t]
    \centering
    \begin{minipage}[t]{0.3\textwidth}
        \centering
        \includegraphics[width=\linewidth]{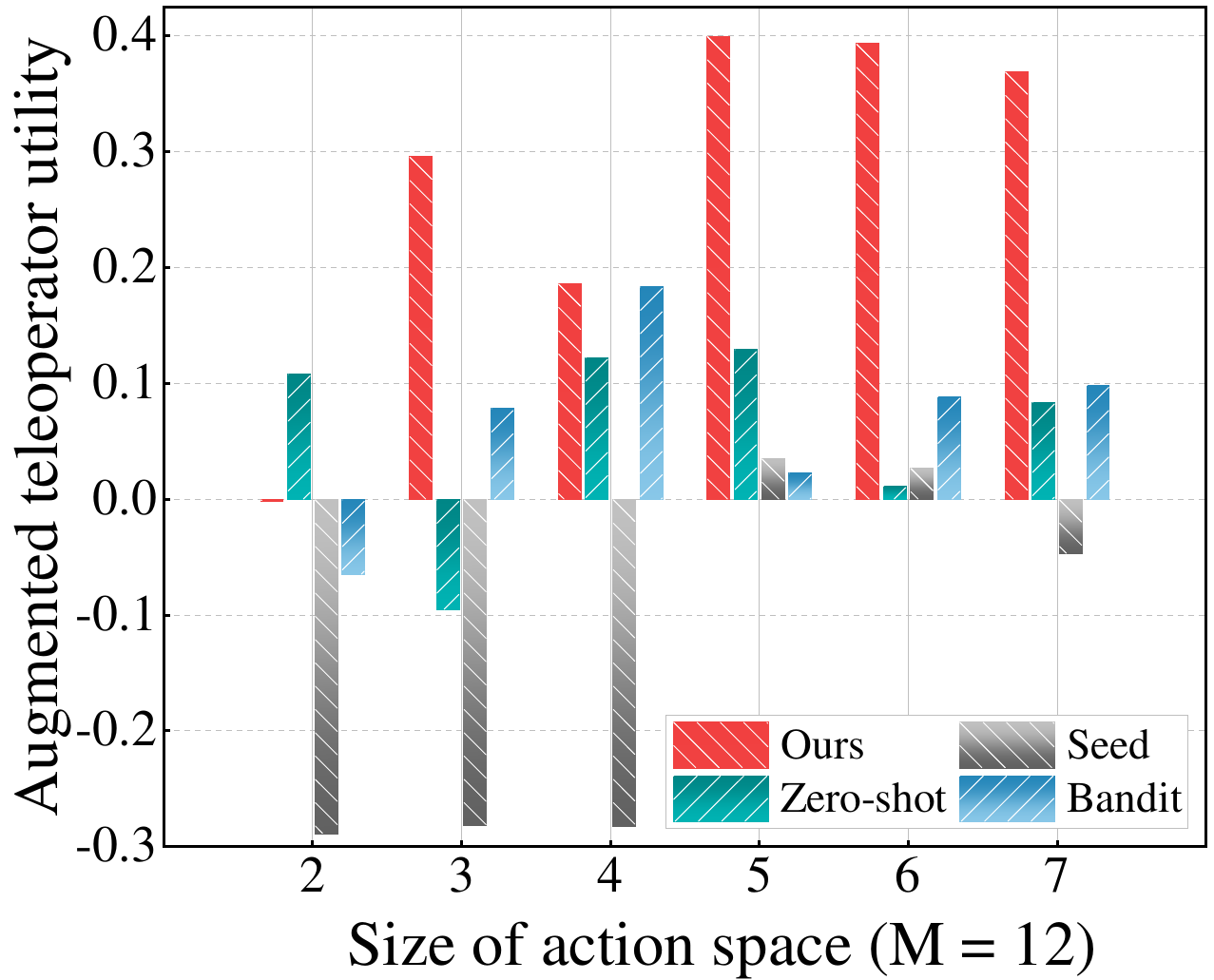}
        \caption{Comparison results by varying the size of the action space $N$ in terms of $\pi_T^{\%}$.}
        \label{fig17}
    \end{minipage}
    \hfill
    \begin{minipage}[t]{0.3\textwidth}
        \centering
        \includegraphics[width=\linewidth]{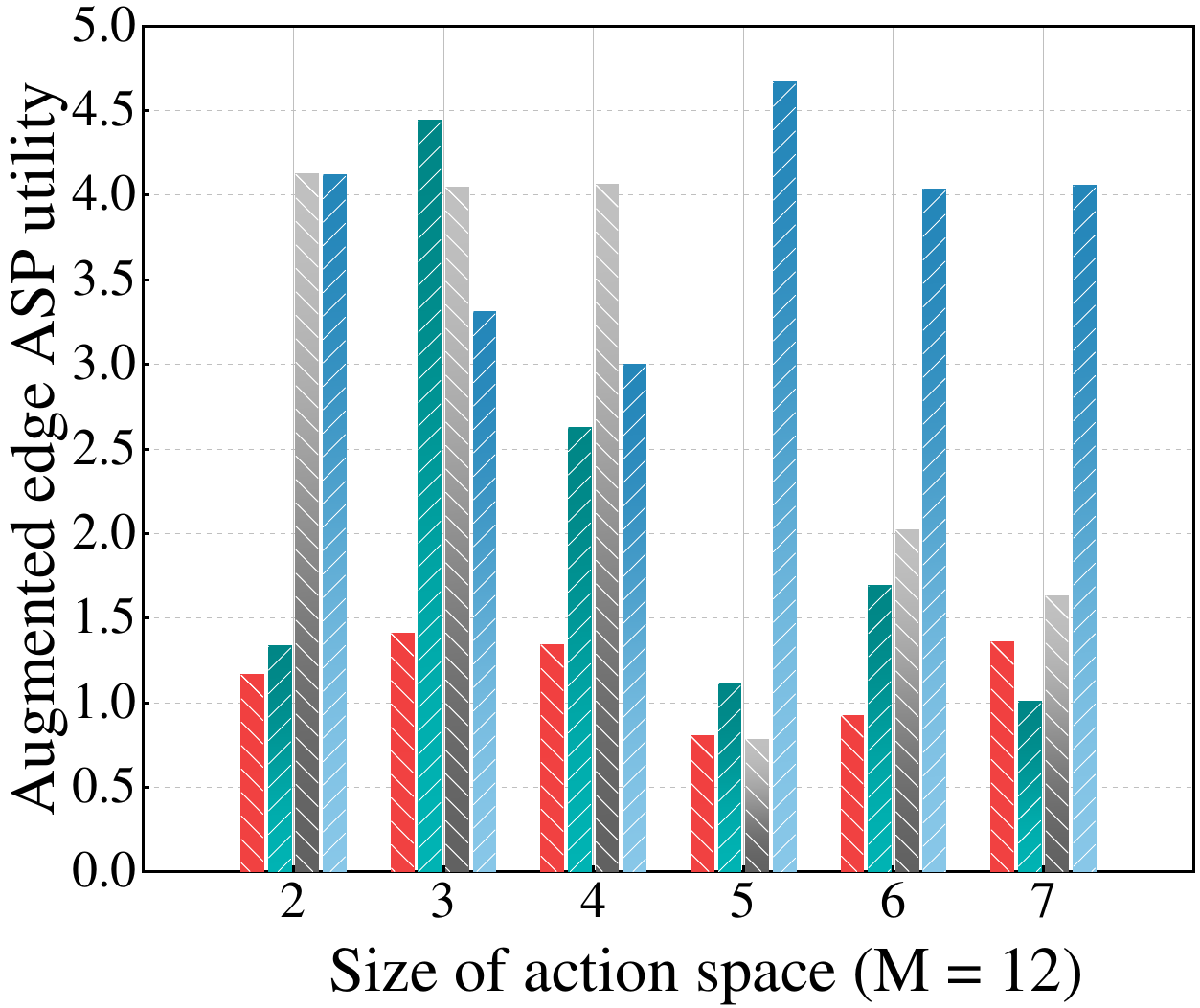}
        \caption{Comparison results by varying the size of the action space $N$ in terms of $\pi_A^{\%}$.}
        \label{fig18}
    \end{minipage}
    \hfill
    \begin{minipage}[t]{0.3\textwidth}
        \centering
        \includegraphics[width=\linewidth]{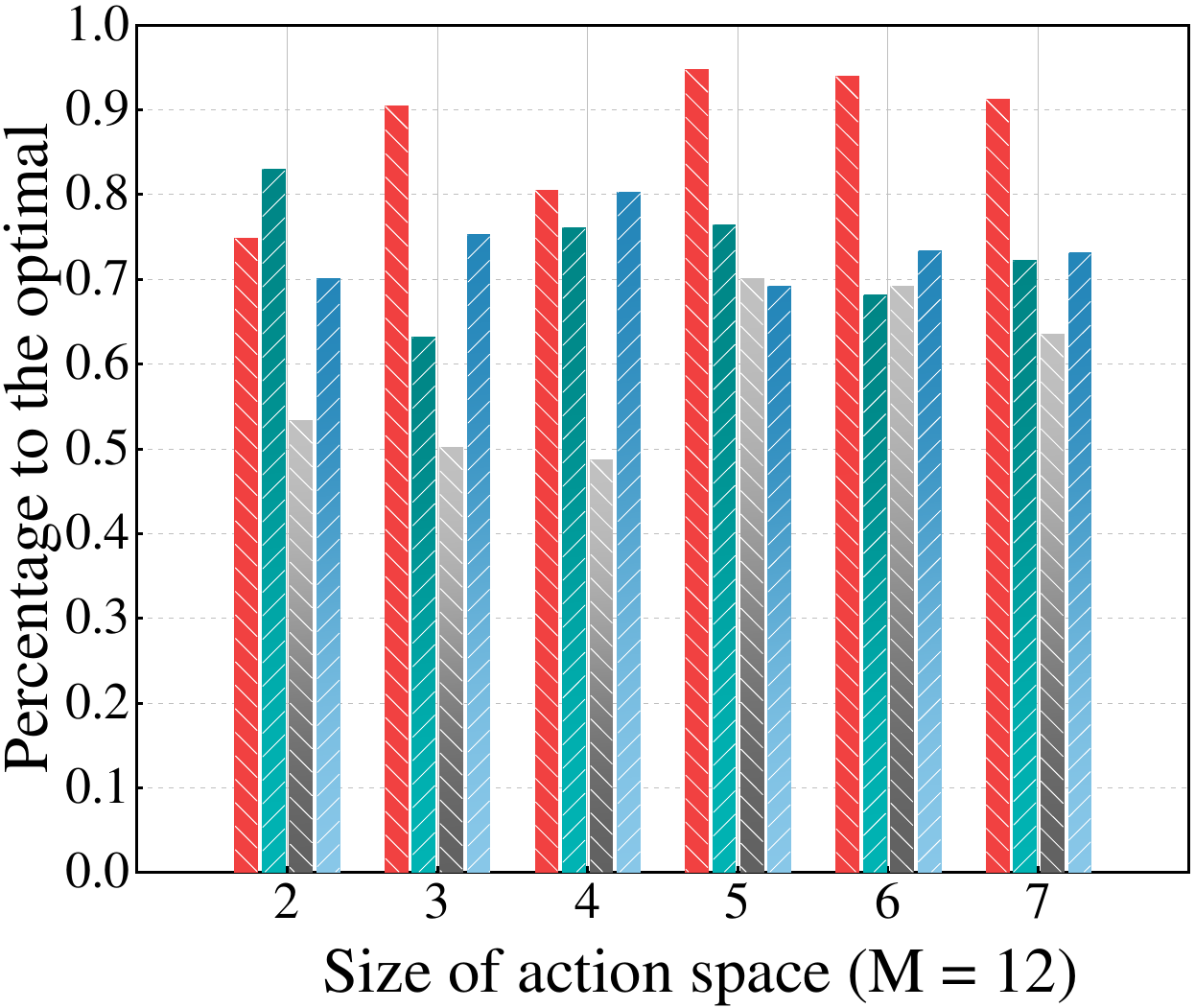}
        \caption{Comparison results by varying the size of the action space $N$ in terms of $\eta$.}
        \label{fig19}
    \end{minipage}
\end{figure*}

The results presented in Figs. \ref{fig14}, \ref{fig15}, and \ref{fig16} is akin to Figs. \ref{fig11}, \ref{fig12}, and \ref{fig13}. Concretely, observing Fig. \ref{fig14}, our proposed method can augment $\pi_T^{\%}$ in around $16 \sim 37\%$ in comparison with the seed solver under varying $M$. Comparing with the bandit algorithm, our proposed method can improve $\pi_T^{\%}$ in around $5 \sim 36 \%$ under varying $M$. Notably, the zero-shot method can achieve the same $\pi_T^{\%}$ as our proposed method when $M=2$. The results in Fig. \ref{fig15} are similar to Fig. \ref{fig12}; we will not reiterate them. From Fig.\ref{fig16}, we observe that our proposed method remains maintain $\eta$ at around $90 \%$ under varying $M$.

Observing Fig. \ref{fig17}, one exception occurs, where our proposed method is inferior to the zero-shot when $N=2$ and only achieves the second-best performance in $\pi_T^{\%}$. However, our proposed method can enlarge the improvement in $\pi_T^{\%}$ to around $35 \sim 58\%$ in comparison with the seed solver under varying $N$. In comparison with the bandit algorithm, our proposed method can improve $\pi_T^{\%}$ by at most $38\%$. Analysis of Fig. \ref{fig18} demonstrates that our proposed method can provide a positive incentive to the edge ASP by satisfying the IR constraint. Lastly, as depicted in Fig. \ref{fig19}, our proposed method can guarantee $\eta$ above $90\%$ under the setting of $N = \{3, 5, 6, 7\}$, which are align with the results in Fig. \ref{fig17}.

In summary, our proposed method can almost surpass benchmarks in terms of $\pi_T^{\%}$ under varying settings of $\alpha$, $M$, and $N$. Our proposed method can also guarantee a positive incentive to the edge ASP under varying settings. Our proposed method can ensure $\eta=90\%$ in most settings, which outperforms benchmarks.

\section{Conclusion} \label{sec:6}
In this paper, regarding the bonus design problem under the hidden action of the edge ASP, we propose an LLM-empowered online learning contract theory to design the bonus (contract). With our proposed method, the teleoperator can design an effective contract with minimal interaction with the edge ASP, thereby eliciting high-quality AIGC services. Specifically, in light of the teleoperator's inability to access the setting (mapping of diffusion steps to the AIGC service quality and the cost of varying diffusion steps) of the edge ASP, we modeled the bonus design as an online learning contract design problem. Next, considering the inherent APX-hard property of the problem, we decomposed the original problem P1 into P2 and P3. Subsequently, due to P2 remaining NP-hard and with uncertain optimization variables, we proposed an LLM-empowered P2 solver to approximate the edge ASP setting. Lastly, we applied the solution from P2 to solve P3 and derive a near-optimal contract directly.

We evaluated the scalability, sensitivity, and effectiveness of our method through simulations.
\begin{enumerate}
    \item Scalability: With an outcome‐space size of 2 and 100 historical logs, performance remained consistent as we increased the ASP’s action‐space size.
    \item Sensitivity: Across mapping coefficients $\alpha$, our method maintained a performance of approximately $\eta = 90\%$. Performance improved with larger action spaces but declined as outcome spaces increased. It was also sensitive to the number of historical logs.
    \item Effectiveness: Compared to benchmarks, we boosted teleoperator utility by $5 \sim 40\%$ and always ensured positive incentives for the ASP. In most settings, we achieved an efficiency rate of $\eta \geq 90\%$, demonstrating robustness and cost efficiency.
\end{enumerate}

\bibliographystyle{IEEEtran}
\bibliography{zhan}

\end{document}


\maketitle

\appendix

\section{Prompts}
In this section, we present all prompts of LLM agents in Fig. 3 of the main file. Remarkably, we extract common prompts of LLM agents in Section \ref{sec:1.6} for clarity.

\subsection{Prompts of Generator LLM} \label{sec:1.1}
\begin{userprompt}[Overall user prompt for generator LLM]
Write a \texttt{\{func\_name\}} function for \texttt{\{problem\_desc\}}:

\medskip
\texttt{\{func\_desc\}}

\medskip
The ‘v’ example is shown as:
\texttt{\{v\}}

\medskip
The ‘content’ example is shown as:
\texttt{\{contract\_logs\}}.

\medskip
\texttt{\{seed\_func\}}

\medskip
Refer to the format of a trivial design above. Be very creative and give \texttt{\{func\_name\}}\_v2.  
Output code only and enclose your code in a Python block:
\verb|```python ...```|.
\end{userprompt}

We designate \textit{func\_name} here as "agent\_solver" in our code. Moreover, 'v' is the contract, and 'content' is the historical contract interaction logs as depicted in Fig. 2 of the main file.

\subsection{Prompts of Short-Term Reflector LLM} \label{sec:1.2}

\begin{userprompt}[Overall user prompt for short-term reflector LLM]
Below are two \{func\_name\} functions for \{problem\_desc\}:

\medskip
\texttt{\{func\_desc\}}

\medskip
You are provided with two code versions below, where the second version performs better than the first one.

\medskip
\textbf{[Worse code]}\\
\texttt{\{worse\_code\}}

\medskip
\textbf{[Better code]}\\
\texttt{\{better\_code\}}

\medskip
You respond with some hints for inferring better agent settings, based on the two code versions and using less than 20 words.
\end{userprompt}

\subsection{Prompts of Crossover LLM} \label{sec:1.3}
\begin{userprompt}[Overall user prompt for crossover LLM]
Write a \{func\_name\} function for \{problem\_desc\}:

\medskip
\texttt{\{func\_desc\}}

\medskip
The ‘v’ example is shown as:
\texttt{\{v\}}

\medskip
The ‘content’ example is shown as:
\texttt{\{contract\_logs\}}.

\medskip
\textbf{[Worse code]}\\
\texttt{\{func\_signature0\}}\\
\texttt{\{worse\_code\}}

\medskip
\textbf{[Better code]}\\
\texttt{\{func\_signature1\}}\\
\texttt{\{better\_code\}}

\medskip
\textbf{[Reflection]}\\
\texttt{\{reflection\}}

\medskip
\textbf{[Improved code]}\\
Please write an improved function \texttt{\{func\_name\}\_v2}, according to the reflection.
\end{userprompt}

Here, the func\_signature is in the format as follows:
\begin{mintedbox}{python}
    def agent_solver_v{version}(v: np.ndarray, content: list[dict]) -> np.ndarray:
\end{mintedbox}

\subsection{Prompts of Long-Term Reflector LLM} \label{sec:1.4}
\begin{userprompt}[Overall user prompt for long-term reflector LLM]
Below is your prior long-term reflection on designing agent setting solver for \{problem\_desc\}:

\texttt{\{prior\_reflection\}}

\medskip
Below are some newly gained insights.

\texttt{\{new\_reflection\}}

\medskip
Write constructive hints for inferring better agent settings, based on prior reflections and new insights, using less than 50 words.
\end{userprompt}

Here, \textit{prior\_reflection} indicates the long-term reflection from the last epoch, and \textit{new\_reflection} is all short-term reflections in this epoch.

\subsection{Prompts of Mutation LLM} \label{sec:1.5}
\begin{userprompt}[Overall user prompt for mutation LLM]
Write a \{func\_name\} function for \{problem\_desc\}:

\medskip
\texttt{\{func\_desc\}}

\medskip
The ‘v’ example is shown as:
\texttt{\{v\}}

\medskip
The ‘content’ example is shown as:
\texttt{\{contract\_logs\}}.

\medskip
\textbf{[Prior reflection]} \\
\texttt{\{reflection\}}

\medskip
\textbf{[Code]}\\
\texttt{\{func\_signature1\}}\\
\texttt{\{elitist\_code\}}

\medskip
\textbf{[Improved code]}\\
Please write a mutated function \texttt{\{func\_name\}\_v2}, according to the reflection.
\end{userprompt}

\subsection{Common Prompts} \label{sec:1.6}

\begin{userprompt}[System prompt for generator LLM]
You are an expert in online learning contract design. Infer a valid agent setting from historical interaction logs to augment the principal’s utility under the agent’s IR and IC constraints. 

\medskip

Output Python code only, formatted as a Python code block: \verb|```python ...```|.
\end{userprompt}

Notably, the generator LLM, crossover LLM, and mutation LLM utilize the same system prompt as shown above.

\begin{userprompt}[System prompt for reflector LLM]
You are an expert in online learning contract design. Your task is to give hints to help infer a better agent setting that not only fits all historical interaction logs but also augments the principal’s utility under the agent’s IR and IC constraints.
\end{userprompt}

The short-term reflector LLM and long-term reflector LLM leverage the same system prompt as shown above.

\begin{userprompt}[Prompt of problem\_desc]
Inferring a valid agent setting via \texttt{agent\_solver} that satisfies all historical interaction logs between the principal and agent in the online contract design problem.
\end{userprompt}

\begin{userprompt}[Prompt of func\_desc]
The \texttt{agent\_solver} function takes a principal’s reward \texttt{v} and historical interaction logs \texttt{content} as inputs.

\medskip
Each log includes:
\begin{itemize}[leftmargin=*]
  \item \textbf{Contract}: a $\{len\_w\}$-dimensional payment vector for $\{len\_w\}$ outcomes;
  \item \textbf{Principal Utility}: the principal’s utility under the contract (zero if the agent rejects);
  \item \textbf{Agent Action}: \texttt{1} for acceptance (expected utility $\ge0$) and \texttt{-1} for rejection (expected utility $<0$).
\end{itemize}

\medskip
The function returns an inferred valid agent setting as an $n \times (\{len\_w\} + 1)$ matrix:
\begin{itemize}[leftmargin=*]
  \item $n$ (number of actions) is chosen to sufficiently explain the data;
  \item Each row corresponds to one possible agent action;
  \item The first $\{len\_w\}$ columns are probabilities over the $\{len\_w\}$ outcomes (summing to 1);
  \item The final column is the nonnegative cost of performing that action.
\end{itemize}
\end{userprompt}

We present the Python code of \textit{seed\_func} as follows:
\begin{mintedbox}{python}
def agent_solver_v1(v: np.ndarray, content: list[dict]) -> np.ndarray:
    n_candidates = 7
    m_outcomes = v.shape[0]
    L = len(content)

    def mini_lp_p(w: np.ndarray, u: float) -> np.ndarray | None:
        """Solve for agent outcome distribution p given wage vector w and utility u."""
        m = len(w)
        A_eq = [np.ones(m), v - w]
        b_eq = [1.0, u]
        bounds = [(0, 1)] * m
        res = linprog(w, A_eq=A_eq, b_eq=b_eq, bounds=bounds, method='highs')
        return res.x if res.success else None

    # Step 1: Filter accepted contracts and solve mini LPs
    candidate_ps = []
    for log in content:
        if log['Agent Action'] == 1:
            w_i = log['Contract']
            u_i = log['Principal Utility']
            p_i = mini_lp_p(w_i, u_i)
            if p_i is not None:
                candidate_ps.append(p_i)

    if not candidate_ps:
        raise ValueError("No valid accepted logs to infer agent strategies.")

    all_p = np.array(candidate_ps)

    # Step 2: Cluster inferred p vectors
    kmeans = KMeans(n_clusters=n_candidates, random_state=0, n_init=10).fit(all_p)
    p0 = kmeans.cluster_centers_

    # Step 3: Assign each accepted log to best-fitting action
    assigns = np.full(L, -1, dtype=int)
    for i, log in enumerate(content):
        if log['Agent Action'] == 1:
            w = log['Contract']
            assigns[i] = int(np.argmax(p0 @ w))

    # Step 4: Compute IR-consistent cost for each action
    c_ir = np.zeros(n_candidates)
    for a in range(n_candidates):
        idx = np.where(assigns == a)[0]
        if idx.size > 0:
            wages = np.array([content[i]['Contract'] for i in idx]).T
            c_ir[a] = p0[a] @ wages.min(axis=1)
        else:
            c_ir[a] = 0.0

    # Step 5: Ensure rejection consistency
    rej_idx = [i for i, log in enumerate(content) if log['Agent Action'] == -1]
    if rej_idx:
        wages_rej = np.array([content[i]['Contract'] for i in rej_idx]).T
        rej_utils = p0 @ wages_rej
        c_rej = rej_utils.max(axis=1)
    else:
        c_rej = np.zeros(n_candidates)

    # Step 6: Final cost = max(IR, rejection threshold)
    c_init = np.maximum(c_ir, c_rej)

    # Step 7: Format agent setting
    agent_setting = np.hstack([p0, c_init[:, np.newaxis]])
    return agent_setting
\end{mintedbox}

We present the Python code of an example worse code as follows:
\begin{mintedbox}{python}
def agent_solver_v1(v: np.ndarray, content: list[dict]) -> np.ndarray:
    """
    Infer a valid agent setting matrix (actions x [12 outcome probs + 1 cost]) that explains
    all historical logs under IR and IC constraints.
    
    Strategy:
    1. Separate accept/reject logs.
    2. For accepted contracts, estimate agent outcome probs and costs via LP:
       - p: probabilities sum to 1
       - cost = p @ wage - agent utility >= 0 and agent utility >= 0 (acceptance)
    3. Cluster these candidate p vectors to represent discrete agent actions.
    4. For each cluster/action, estimate minimal cost consistent with logs.
    5. For rejected contracts, enforce that for all actions, expected utility < 0.
    6. Adjust costs upward if needed to maintain rejection consistency.
    7. Return matrix with rows: [p (12), cost (>=0)]
    
    Args:
        v (np.ndarray): Principal reward vector of length 12.
        content (list[dict]): List of logs with keys 'Contract' (12-vector), 
                              'Principal Utility' (float), 'Agent Action' (1 or -1).
    
    Returns:
        np.ndarray: n_actions x 13 matrix (12 probs + 1 cost).
    """
    m_outcomes = v.shape[0]
    logs = pd.DataFrame(content)
    L = len(logs)
    
    # Separate accepted and rejected logs
    accepted = logs[logs['Agent Action'] == 1]
    rejected = logs[logs['Agent Action'] == -1]
    
    # Step 1: For each accepted log, solve LP to find p and agent utility >= 0
    # Objective: minimize ||p - w||_2 (heuristic) subject to:
    #   sum(p) = 1
    #   p @ wage - cost >= 0 (agent utility >= 0)
    #   cost = minimum agent cost
    # Instead of complicated QP, simplify: assume agent utility = 0 at acceptance (IR tight),
    # so cost = p @ wage, find p minimizing ||p - normalized wage||_2 to guess p.
    
    def lp_find_p_and_cost(w):
        # Solve for p: sum p =1, p >=0, cost = p @ w
        # LP: minimize sum |p - w_norm| (approximate by linprog twice)
        # Here, just take p proportional to wage (w+eps), normalized.
        w = np.array(w)
        w = np.clip(w, 0, None)
        if w.sum() == 0:
            p = np.ones(m_outcomes) / m_outcomes
        else:
            p = w / w.sum()
        cost = p @ w  # agent utility 0 => cost = expected wage
        return p, cost
    
    candidate_p = []
    candidate_cost = []
    for _, row in accepted.iterrows():
        w = row['Contract']
        p_i, cost_i = lp_find_p_and_cost(w)
        candidate_p.append(p_i)
        candidate_cost.append(cost_i)
    candidate_p = np.array(candidate_p)
    candidate_cost = np.array(candidate_cost)
    
    # Step 2: Cluster accepted p vectors into discrete agent actions
    # Use AgglomerativeClustering with unknown n_clusters, select by distance threshold
    # To be adaptive, try cluster numbers from 2 to 10 and pick best silhouette-like measure
    from sklearn.metrics import silhouette_score
    
    best_n = 1
    best_score = -1
    best_labels = None
    for n_clusters in range(2, min(11, len(candidate_p)+1)):
        clustering = AgglomerativeClustering(n_clusters=n_clusters, linkage='ward').fit(candidate_p)
        try:
            score = silhouette_score(candidate_p, clustering.labels_)
            if score > best_score:
                best_score = score
                best_n = n_clusters
                best_labels = clustering.labels_
        except:
            continue
    if best_labels is None:
        best_n = 1
        best_labels = np.zeros(len(candidate_p), dtype=int)
    
    # Compute cluster centers (mean p) and minimal costs per cluster
    p_actions = np.zeros((best_n, m_outcomes))
    c_actions = np.zeros(best_n)
    for a in range(best_n):
        idx = np.where(best_labels == a)[0]
        if len(idx) == 0:
            # Fallback uniform
            p_actions[a] = np.ones(m_outcomes) / m_outcomes
            c_actions[a] = 0.0
            continue
        p_actions[a] = candidate_p[idx].mean(axis=0)
        # Cost is minimal over cluster
        c_actions[a] = candidate_cost[idx].min()
    
    # Normalize p-actions to sum to 1 (numerical safety)
    p_actions = np.clip(p_actions, 0, None)
    p_actions = p_actions / p_actions.sum(axis=1, keepdims=True)
    
    # Step 3: Enforce IR: agent utility = p @ wage - cost >= 0 for all accepted logs assigned to action
    # Assign each accepted log to best action (max p @ wage - cost)
    a_assign = np.zeros(len(accepted), dtype=int)
    accepted_wages = np.stack(accepted['Contract'].to_numpy())
    for i, w in enumerate(accepted_wages):
        utilities = p_actions @ w - c_actions
        a_assign[i] = np.argmax(utilities)
    
    # Adjust costs upward to satisfy IR for assigned accepted logs
    for a in range(best_n):
        idx = np.where(a_assign == a)[0]
        if len(idx) == 0:
            # No logs assigned, keep cost as is
            continue
        w_assigned = accepted_wages[idx]
        u = p_actions[a] @ w_assigned.T - c_actions[a]
        min_u = u.min()
        if min_u < 0:
            c_actions[a] -= min_u  # raise cost to fix IR
    
    # Step 4: Enforce IC on rejected logs: for all actions utility < 0
    if len(rejected) > 0:
        rejected_wages = np.stack(rejected['Contract'].to_numpy())  # shape R x 12
        for a in range(best_n):
            utilities_rej = p_actions[a] @ rejected_wages.T - c_actions[a]  # shape R
            max_u_rej = utilities_rej.max()
            if max_u_rej >= 0:
                # Increase cost to make all rejected utilities < 0 by a margin (e.g., 1e-8)
                c_actions[a] += max_u_rej + 1e-8
    
    # Step 5: Ensure non-negative costs
    c_actions = np.clip(c_actions, 0, None)
    
    # Step 6: Final sanity check: probs sum to 1, non-negative
    p_actions = np.clip(p_actions, 0, None)
    p_actions = p_actions / p_actions.sum(axis=1, keepdims=True)
    
    # Compose final agent setting
    agent_setting = np.hstack([p_actions, c_actions[:, None]])
    return agent_setting
\end{mintedbox}

We present the Python code of an example better code as follows:
\begin{mintedbox}{python}
def agent_solver_v1(v: np.ndarray, content: list[dict]) -> np.ndarray:
    """
    Infer a valid agent setting from historical logs by clustering accepted contracts' implied
    outcome distributions and estimating agent costs to satisfy IR and IC constraints.
    This version uses DBSCAN clustering for adaptive action number and robust cost inference.
    """
    m_outcomes = v.shape[0]
    L = len(content)
    
    # Separate accepted and rejected logs
    accepted_logs = [log for log in content if log['Agent Action'] == 1]
    rejected_logs = [log for log in content if log['Agent Action'] == -1]
    
    if len(accepted_logs) == 0:
        raise ValueError("No accepted contracts; cannot infer agent strategies.")
    
    # Step 1: For each accepted contract, infer an agent outcome distribution p by solving LP:
    # maximize p @ w subject to p sums to 1 and expected utility = principal utility (approximate)
    # Instead of exact equality, allow small tolerance to handle noise.
    
    def infer_p(w, u, tol=1e-6):
        # Maximize expected wage p @ w s.t sum p=1, p>=0, and p@(v - w) = 0
        # We relax equality to two inequalities to handle numeric instability:
        # u - tol <= p@(v - w) <= u + tol
        c = -np.array(w)  # maximize p@w -> minimize -p@w
        A_eq = [np.ones(m_outcomes)]
        b_eq = [1.0]
        A_ub = [np.array(v) - np.array(w), -(np.array(v) - np.array(w))]
        b_ub = [u + tol, -u + tol]
        bounds = [(0, 1)] * m_outcomes
    
        res = linprog(c, A_ub=A_ub, b_ub=b_ub, A_eq=A_eq, b_eq=b_eq, bounds=bounds, method='highs')
        if res.success:
            p = res.x
            p[p < 0] = 0
            p = p / p.sum()  # Normalize in case of slight numerical issues
            return p
        else:
            return None
    
    ps = []
    us = []
    for log in accepted_logs:
        w = np.array(log['Contract'])
        u = np.maximum(np.dot(v, np.ones(m_outcomes)) - log['Principal Utility'], 0)
        # Actually, agent utility = p @ w - cost, but cost unknown.
        # Use agent IR: expected utility >= 0
        # We approximate agent utility by enforcing p@(w - v) >= 0 as proxy.
        # Here set u=0 and just maximize p@w with p@1=1
        p = infer_p(w, 0.0)
        if p is not None:
            ps.append(p)
            us.append(log['Principal Utility'])
    if len(ps) == 0:
        raise ValueError("Failed to infer distributions for accepted contracts.")
    
    ps = np.array(ps)
    
    # Step 2: Cluster inferred p to find distinct agent actions adaptively using DBSCAN
    # DBSCAN can detect number of clusters automatically and handle noise
    # Use cosine metric to cluster outcome distributions
    from sklearn.metrics.pairwise import cosine_distances
    
    dist_mat = cosine_distances(ps)
    # eps chosen small to cluster similar ps; min_samples=2 to form clusters
    clustering = DBSCAN(eps=0.05, min_samples=2, metric='precomputed').fit(dist_mat)
    labels = clustering.labels_
    
    # Handle noise points by assigning them as separate clusters
    unique_labels = set(labels)
    if -1 in unique_labels:
        noise_count = np.sum(labels == -1)
        max_label = max(unique_labels)
        for i, lab in enumerate(labels):
            if lab == -1:
                max_label += 1
                labels[i] = max_label
        unique_labels = set(labels)
    
    n_actions = len(unique_labels)
    
    # Step 3: Compute cluster centers as representative outcome distributions for each action
    p_actions = np.zeros((n_actions, m_outcomes))
    for action in unique_labels:
        idx = np.where(labels == action)[0]
        p_actions[action] = ps[idx].mean(axis=0)
        p_actions[action] /= p_actions[action].sum()  # Normalize
    
    # Step 4: Estimate costs for each inferred action to satisfy IR (accepted) and IC (rejected)
    
    # For accepted logs, assign them to nearest action by cosine similarity on p
    def nearest_action(p):
        sims = p_actions @ p
        return np.argmax(sims)
    
    accepted_indices = [i for i, log in enumerate(content) if log['Agent Action'] == 1]
    assigned_actions = np.zeros(len(accepted_indices), dtype=int)
    for i, idx in enumerate(accepted_indices):
        contract = np.array(content[idx]['Contract'])
        # Find expected utility under each p_action: p@w - cost (cost unknown)
        # We assign by max dot product p@w, approximated by p_action@contract
        dots = p_actions @ contract
        assigned_actions[i] = np.argmax(dots)
    
    # For each action, IR cost <= min over accepted logs assigned of p_action@w
    c_ir = np.full(n_actions, np.inf)
    for a in range(n_actions):
        assigned_idx = [accepted_indices[i] for i in range(len(assigned_actions)) if assigned_actions[i] == a]
        if len(assigned_idx) > 0:
            costs = []
            for idx in assigned_idx:
                contract = np.array(content[idx]['Contract'])
                # Agent utility = p@w - cost >= 0 => cost <= p@w
                costs.append(p_actions[a] @ contract)
            c_ir[a] = min(costs)
        else:
            c_ir[a] = 0.0  # No assigned accepted contract, cost zero
    
    # For rejected logs, ensure no action yields agent utility >= 0,
    # i.e. for all a and rejected contract w_rej: p_action@w_rej - cost < 0 => cost > p_action@w_rej
    if len(rejected_logs) > 0:
        for a in range(n_actions):
            max_rej_utility = -np.inf
            for log in rejected_logs:
                w_rej = np.array(log['Contract'])
                util = p_actions[a] @ w_rej
                if util > max_rej_utility:
                    max_rej_utility = util
            # Cost must be strictly greater than max_rej_utility, so at least max_rej_utility + eps
            c_ir[a] = max(c_ir[a], max_rej_utility + 1e-8)
    else:
        # No rejection constraints
        pass
    
    # Final costs are non-negative
    c_ir = np.maximum(c_ir, 0)
    
    # Step 5: Return agent setting matrix: rows = actions, columns = 12 probs + 1 cost
    agent_setting = np.hstack([p_actions, c_ir[:, None]])
    
    return agent_setting
\end{mintedbox}